\documentclass[draftcls, onecolumn]{IEEEtran}
\usepackage[T1]{fontenc}
\usepackage{verbatim}
\usepackage{amsmath}
\usepackage{amsthm}
\usepackage{amssymb}
\usepackage{graphicx}
\PassOptionsToPackage{normalem}{ulem}
\usepackage{ulem}

\makeatletter
\theoremstyle{plain}
\newtheorem{thm}{\protect\theoremname}
\theoremstyle{definition}
\newtheorem{defn}[thm]{\protect\definitionname}
\theoremstyle{plain}
\newtheorem{fact}[thm]{\protect\factname}
\theoremstyle{plain}
\newtheorem{prop}[thm]{\protect\propositionname}

\usepackage{changepage}

\newcommand{\var}{\mathrm{var}}

\IEEEoverridecommandlockouts

\newcommand{\bmat}{\left[\begin{array}}
\newcommand{\emat}{\end{array}\right]}

\makeatother

\providecommand{\definitionname}{Definition}
\providecommand{\factname}{Fact}
\providecommand{\propositionname}{Proposition}
\providecommand{\theoremname}{Theorem}

\begin{document}

\title{Data Discovery and Anomaly Detection Using Atypicality: Theory}

\author{Anders H{\o}st-Madsen, \emph{Fellow, IEEE}, Elyas Sabeti, \emph{Member,
IEEE}, Chad Walton\thanks{A. H{\o}st-Madsen and E. Sabeti are with the Department of Electrical
Engineering, University of Hawaii Manoa, Honolulu, HI 96822 (e-mail:
\{ahm,sabeti\}@hawaii.edu). C. Walton is with the Department of Surgery,
University of Hawaii, Honolulu, HI, 96813. Email: cwalton@hawaii.edu.
This work was supported in part by NSF grants CCF 1017823, 1017775,
and\textbf{ }1434600. The paper was presented in part at IEEE Information
Theory Workshop 2013, Seville.}}

\maketitle
\global\long\def\cov{\mathrm{cov}}

\begin{abstract}
A central question in the era of 'big data' is what to do with the
enormous amount of information. One possibility is to characterize
it through statistics, e.g., averages, or classify it using machine
learning, in order to understand the general structure of the overall
data. The perspective in this paper is the opposite, namely that most
of the value in the information in some applications is in the parts
that deviate from the average, that are unusual, atypical. We define
what we mean by 'atypical' in an axiomatic way as data that can be
encoded with fewer bits in itself rather than using the code for the
typical data. We show that this definition has good theoretical properties.
We then develop an implementation based on universal source coding,
and apply this to a number of real world data sets.
\end{abstract}


\section{Introduction}

One characteristic of the information age is the exponential growth
of information, and the ready availability of this information through
networks, including the internet \textendash{} ``Big Data.'' The
question is what to do with this enormous amount of information. One
possibility is to characterize it through statistics \textendash{}
think averages. The perspective in this paper is the opposite, namely
that most of the value in the information is in the parts that deviate
from the average, that are unusual, atypical. The rest is just background
noise. Take art: the truly valuable paintings are those that are rare
and atypical. The same could be true for scientific research and entrepreneurship.
Take online collections of photos, such as Flickr.com. Most of the
photos are rather pedestrian snapshots and not of interest to a wider
audience. The photos that of interest are those that are unique. Flickr
has a collection of photos rated for 'interestingness,' and one can
notice that those photos are indeed very different from typical photos.
They are atypical. 

The aim of our approach is to extract such 'rare interesting' data
out of big data sets. The central question is what 'interesting' means.
A first thought is to focus on the 'rare' part. That is, interesting
data is something that is unlikely based on prior knowledge of typical
data or examples of typical data, i.e., training. This is the way
an outlier is usually defined. Unlikeliness could be measured in terms
of likelihood, in terms of codelength \cite{AkogluAl12,SmetsVreeken11}
\textendash{} called 'surprise' in \cite{LiuPrincipe09} \textendash{}
or according to some distance measure. This is also the most common
principle in anomaly detection \cite{ChandolaAl12}. However, perhaps
being unlikely is not sufficient for something to be 'interesting.'
In many cases, outliers are junk that are eliminated not to contaminate
the typical data. What makes something interesting is maybe that it
has a new unusual structure in itself that is quite different from
the structure of the data we have already seen. Return to the example
of paintings: what make masterworks interesting is not just that they
are different than other paintings, but that they have some 'structure'
that is intriguing. Or take another example. Many scientific discoveries,
like the theory of relativity and quantum mechanics, began with experiments
that did not fit with prevailing theories. The experiments were outliers
or anomalies. What made them truly interesting was that it was possible
to find a new theory to explain the data, be it relativity or quantum
mechanics. This is the principle we pursue: finding data that have
better alternative explanations than those that fit the typical data.

Something being unlikely is not even necessary for the data to be
'interesting.' Suppose the typical data is iid uniform $\{0,1\}$.
Then any sequence of bits are equally likely. Therefore, a sequence
consisting of purely 1, $1111111\ldots$ is in no way 'surprising.'
Yet, it should catch our interest.

When we look for new interesting data, a characteristic is that we
do not know what we are looking for. We are looking for ``unknown
unknowns'' \cite{Rumsfeld}. Instead of looking at specific statistics
of data, we need to use a universal approach. This is provided by
information theory. 

This idea of finding alternative explanations for data rather than
measuring some kind of difference from typical data is what separates
our method from usual approaches in outlier detection and anomaly
detection. As far as we can determine from reading hundreds of papers,
our approach has not been explored previously. Obviously, information
theory and coding have been used in anomaly detection, data mining,
and knowledge discovery before, and we will discuss how this compares
to our approach later. Our methodology also has connections to tests
for randomness, e.g., the run length test and \cite{HamanoYamamoto08,NiesBook},
but our aim is different.  

\subsection{\label{Applications.sec}Applications}

Atypicality is relevant in large number of various applications. We
will list a few applications here. 

\medskip{}

\noindent \textbf{ECG. }For electrocardiogram (ECG) recordings there
are patterns in heart rate variability that are known to indicate
possible heart disease \cite{Malik96,M.F.Hilton:1999fj,N.V.Thakor:1991gf,ThayerAl09}.
With modern technology it is possible for an individual to wear and
unobtrusive heart rate monitor 24/7. If atypical patterns occur, it
could be indicative of disease, and the individual or a doctor could
be notified. But perhaps a more important application is to medical
research. One can analyze a large collection of ECG recordings and
look for individuals with atypical patterns. This can then potentially
be used to develop new diagnostic tools. 

\noindent \textbf{Genomics}. Another example of application is interpretation
of large collections of genomics data. Given that all mammals have
essentially the same set of genes, there must exist some significant
differences that distinguish the obvious distinct attributes between
species, as well as more subtle differences within a species. Although
the genome has been mined by exhaustive studies applying a panoply
of approaches, regions once thought to be \textquotedblleft uninteresting\textquotedblright{}
have recently come under increased study for their potential role
in defined morphological and physiological differences between individuals
\cite{FondonGarner04}. Applying an atypical evaluation tool to genomic
data from individuals of known pathophysiological/morphological irregularities
may provide valuable insight to the genetic mechanisms underlying
the condition. 

\noindent \textbf{Ocean Monitoring}. In passive acoustic monitoring
(PAM) \cite{Mellinger07} of oceans, one or more hydrophones is towed
behind a ship or deployed in a fixed bottom-mounted or suspended array
in order to record vocalizations of marine mammals. One major focus
is to detect, and perhaps count, rare or endangered species. It would
be highly interesting to scan the data for any unusual patterns, which
can then be further examined by a researcher. 

\noindent \textbf{Plant Monitoring}. In for example nuclear plants,
atypical monitoring data may be indicative of something about to go
wrong.

\noindent \textbf{Computer Networks}. Atypical network traffic could
be indicative of a cyberattack. This is already being used through
anomaly detection \cite{ThottanJi03}. However, an abstract atypicality
approach can be used to find more subtle attacks \textendash{} the
unknown unknowns.

\noindent \textbf{Airport Security}. Already software is being used
to flag suspicious flyers, likely based on past attacks. Atypical
detection could be used to find innovative attackers.

\noindent \textbf{Stock Market}. Atypicality could be used to detect
insider trading. It could also be used by investors to find unusual
stocks to invest in, promising outstanding returns \textendash{} or
ruin.

\noindent \textbf{Astronomy}. Atypicality can be used to scan huge
databases for new kinds of cosmological phenomena.%

\noindent \textbf{Credit Card Fraud}. Unusual spending patterns could
be indicative of fraud. This is already used by credit card companies,
but obviously in a simple, and annoying way, as anyone who's credit
card has been blocked on an overseas trip can testify to.

\noindent \textbf{Gambling}. Casinos are constantly fighting fraudsters.
This is a game of cat and mouse. Fraudsters constantly find new ways
to trick the casinos (one such inventor was Shannon himself). Therefore,
an abstract atypicality approach may be the best solution to catch
new ways of fraud.

\subsection{Notation}

We use $x$ to denote a sequence in general, and $x^{l}$ when we
need to make the length explicit; $x_{i}$ denotes a single sample
of the sequence. We use capital letters $X_{i}$ to denote random
variables rather than specific outcomes. Finally $\mathcal{X}$ denotes
a subsequence. All logarithms are to base 2 unless otherwise indicated.

\section{\label{MDL.sec}Atypicality}

Our starting point is the in theory of randomness developed by Kolmogorov
and Martin-L\"{o}f \cite{LiVitanyi,NiesBook,CoverBook}. Kolmogorov
divides (infinite) sequences into 'typical' and 'special.' The typical
sequences are those that we can call random, that is, they satisfy
all laws of probability. They can be characterized through Kolmogorov
complexity. A sequence of bits $\{x_{n},n=1,\ldots,\infty\}$ is random
(i.e, iid uniform) if the Kolmogorov complexity of the sequence satisfies
$K(x_{1},\ldots,x_{n})\geq n-c$ for some constant $c$ and for all
$n$\cite{LiVitanyi}. The sequence is incompressible if $K(x_{1},\ldots,x_{n}|n)\geq n$
for all $n$, and a finite sequence is algorithmically random if $K(x_{1},\ldots,x_{n}|n)\geq n$
\cite{CoverBook}. In terms of coding, an iid random sequence is also
incompressible, or, put another way, the best coder is the identity
function. Let us assume we draw sequences $x^{n}$ from an iid uniform
distribution. The optimum coder is the identity function, and the
code length is $n$. Now suppose that for one of these sequences we
can find a (universal) coder so that the code length is less than
$n$; while not directly equivalent, one could state this as $K(x_{1},\ldots,x_{n}|n)<n$.
With an interpretation of Kolmogorov's terms, this would not be a
'typical' sequence, but a 'special' sequence. We will instead call
such sequences 'atypical.' Considering general distributions and general
(finite) alphabets instead of iid uniform distributions, we can state
this in the following general principle\begin{adjustwidth}{0.25in}{0.25in}
\begin{defn}
\label{atypdef.thm}\emph{A sequence is atypical if it can be described
(coded) with fewer bits in itself rather than using the (optimum)
code for typical sequences}.
\end{defn}
\end{adjustwidth}\medskip{}

This definition is central to our approach to the atypicality problem.

In the definition, the ``(optimum) code for typical sequences,''
is quite specific, following the principles in for example \cite{CoverBook}.
We assume prefix free codes. Within that class the coding could be
done using Huffman codes, Shannon codes, Shannon-Fano-Elias codes,
arithmetic coding etc. We care only about the code length, and among
these the variation in length is within a few bits, so that the code
length for typical encoding can be quite accurately calculated.

On the other hand, ``described (coded) with fewer bits in itself''
is less precise. In principle one could use Kolmogorov complexity,
but Kolmogorov complexity is not calculable and it is only given except
for a constant, and comparison with code length therefore is not an
``apples-to-apples'' comparison. Rather, some type of universal
source coder should be used. This can be given a quite precise meaning
in the class of finite state machine sources, \cite{Rissanen86b}
and following work, and is strongly related to minimum description
length (MDL) \cite{Rissanen83,Rissanen84,Rissanen86,Rissanen86b}.
What is essential is that we adhere to \emph{strict decodability}
at the decoder. The decoder only sees a stream of bits, and from this
it should be able to accurately reconstruct the source sequence. So,
for example, if a sequence is atypical, there must be a type of ``header''
telling the decoder to use a universal decoder rather than the typical
decoder. Or, if atypical sequences can be encoded in multiple ways,
the decoder must be informed through the sequence of bits which encoder
was used. One could argue that such things are irrelevant for for
example anomaly detection, since we are not actually encoding sequences.
The problem is that if such terms are omitted, it is far too easy
to encode a sequence ``in itself.'' This is like choosing a more
complex model to fit data, without accounting for the model complexity
in itself, which is exactly what MDL sets out to solve, although also
in this case actual encoding is not done. We therefore try to account
for all factors needed to describe data, and we believe this is one
of the key strengths of the approach.

A major difference between atypical data and anomalous data is that
atypicality is an axiomatic property of data, defined by Definition
\ref{atypdef.thm} based on Kolmogorov-Martin-L\"{o}f randomness.
On the other hand, as far as we know, an anomaly is not something
that can be strictly defined. Usually, we think of an anomaly as something
caused by an outside phenomenon: an intruder in a computer network,
a heart failure, a gambler playing tricks. This influences how we
think of performance. If a detector fails to give an indication of
an anomaly, we have a miss (or type II error), but if it gives an
indication when such things are not happening we have a false alarm
(type I error). Atypicality, on the other hand, is purely a property
of data. Ideally, there are therefore no misses or false alarms: data
is atypical or not. Here is what we mean. If there is an anomaly that
expresses itself through the observed data, that must mean that there
is some structure in the data, and in theory a source coder would
discover and exploit such structure and reduce code length. Thus,
if the data is not atypical that means there is simply no way to detect
the anomaly through the observations \textendash{} again in theory.
We therefore cannot really call that a miss. On the other hand, suppose
that in a casino a gambler has a long sequence of wins. This could
be due to fraud, but it could also be simply due to randomness. But
casino security would be interested in either case for further scrutiny.
Thus, the reason for the atypicality does not really matter, the atypicality
itself matters. Still, to distinguish the two cases we call a sequence
\emph{\uline{intrinsically}} atypical if it is atypical according
to Definition \ref{atypdef.thm} while being generated from the typical
probability model, while it is \emph{\uline{extrinsically}} atypical
if it is in fact generated by any other probability law.

Definition \ref{atypdef.thm} has two parts that work in concert,
and we can write it simplified as $C_{t}(x)-C_{a}(x)>0$ where $C_{t}$
is the typical codelength and $C_{a}$ the atypical codelength. The
typical code length $C_{t}(x)$ is simply an expression of the likelihood
of seeing a particular sequence. If $C_{t}(x)$ is large it means
that the given sequence is unlikely to happen, and detecting sequences
by $C_{t}(x)>\tau$ would catch many outliers. As an extreme example,
if a sequence is impossible according to the typical distribution,
$C_{t}(x)=\infty$, and it would always be caught. But it would not
work universally. If, as we started out with, typical sequences are
iid uniform, any sequence is equally likely and $C_{t}(x)>\tau$ would
not catch any sequences. In this case, if a test sequence has some
structure, it is possible that $C_{a}(x)<C_{t}(x)$, and such sequences
would be caught by atypicality; thus calculating $C_{a}(x)$ is essential.
Calculating $C_{t}(x)$ is also essential. Suppose that we instead
use $E[C_{t}]-C_{a}(x)$, where $E[C_{t}]$ is the code length used
to encode typical sequences ``on average,'' essentially the entropy
rate. Again, this will catch some sequences: if a test sequence has
more or less structure than typical sequences, $E[C_{t}]-C_{a}(x)\neq0$.
But again, it will omit very obvious examples: if as test sequence
we use a typical sequence with $0$ and $1$ swapped, $E[C_{t}]\approx C_{a}(x)$,
while on the other hand $C_{t}(x)>C_{a}(x)$. And impossible sequences
with $C_{t}(x)=\infty$ would not be caught with absolute certainty.
Now, to declare something an outlier, we have to find a coder with
$C_{a}(x)<C_{t}(x)$. It is not sufficient that $C_{t}(x)$ is large,
i.e., that the sequence is unlikely to happen. However, we can always
use the trivial coder that transmits data uncoded. If the sequence
is unlikely to happen according to the typical distribution, then
it is likely that $C_{t}(x)>(\mbox{length of }x)$.

Thus, it can be seen that the two parts work in concert to catch sequences.
Each part might catch some sequences, but to catch all ``anomalies,''
both parts have to be used.

Another point of view is the following. Suppose again the typical
model is binary uniform iid. We look at a collection of sequences,
and now we want to find the \emph{most} atypical sequences, i.e.,
the most ``interesting'' sequences. Without a specification of what
``interesting'' is, it seems reasonable to choose those sequences
that have the most structure, and again this can reasonably be measured
by how much the sequence can be compressed. This is what Rissanen
\cite{Rissanen86b} calls ``useful information,'' $U(x)=n-C_{a}(x)$.
But again, we need to take into account the typical model if it is
not uniform iid. For example, if typical sequences have much structure,
then sequence with little structure might be more interesting. We
therefore end up with that $C_{t}(x)-C_{a}(x)$ is a reasonable measure
of how interesting sequences might be. 

\subsection{\label{Alternatives.sec}Alternative approaches}

While, as argued in the introduction, and outlined above, what we
are aiming for is not anomaly detection in the traditional sense,
there are still many similarities. And certainly information theory
and universal source coding has been used previously in anomaly detection,
e.g., \cite{ChandolaAl12,EvansBarnettAl04,WangAl12,LeeXiang01,PaschalidisSmaragdakis09,HanChoi09,BeligaLin05,ShahriarZulkernine12,XiangLiZhou11,PanWang06,EllandLiebrock06}.
The approaches have mostly been heuristic. A more fundamental and
systematic approach is Information Distance defined in \cite{LiVitanyiAl04}.
Without being able to claim that this applies to \emph{all} of the
perhaps hundreds of papers, we think the various approaches can be
summarized as using universal source coding as a type of distance
measure, whether it satisfies strict mathematical metric properties
as in \cite{LiVitanyiAl04} or is more heuristic. On the other hand,
our methodology in Definition \ref{atypdef.thm} cannot be classified
as a distance measure in a traditional sense. We are instead trying
to find alternative explanations for data. We will comment on how
our approach contrasts with a few other approaches. 

While the similarity distance developed in \cite{LiVitanyiAl04} is
not directly applicable to the problem we consider, we can to some
extent adapt it, which is useful for contrast. The similarity distance
is 
\[
d=\frac{\min\{K(y|x^{*}),K(x|y^{*})\}}{\max\{K(x),K(y)\}}
\]
Instead of being given the typical distribution, we can imagine that
we are given a very long typical sequence $x$ which is used for ``training.''
In that case 
\[
d=\frac{K(x|y^{*})}{K(x)}=\frac{K(x,y)-K(y)}{K(x)}
\]
within a certain approximation. Suppose, as was our starting point
above, that the typical distribution is binary iid uniform. If $y$
is also binary iid uniform, within a constant $K(x,y)=K(x)+K(y)$,
and $d=1$. But if $y$ is drawn from some other distribution, $x$
cannot help describing $x$ either, and still $d=1$. That makes sense:
two completely random sequences are not similar, whether they are
from the same distribution or not. Thus, similarity distance cannot
be used for 'anomaly' detection as we have have defined it: looking
for 'special' sequences in the words of Kolmogorov. This is not a
problem of the similarity metric; it does exactly what it is designed
for, which is really deterministic similarity between sequences, appropriate
for classification. The reason similarity distance still gives results
for anomaly detection \cite{KeoghAl04} is actually that universal
source coders approximate Kolmogorov complexity poorly.

Heuristic methods using for anomaly detection using universal source
coding \cite{ChandolaAl12,EvansBarnettAl04,WangAl12,LeeXiang01,PaschalidisSmaragdakis09,HanChoi09,BeligaLin05,ShahriarZulkernine12,XiangLiZhou11,PanWang06,EllandLiebrock06,AkogluAl12,SmetsVreeken11}
are mostly based on comparing code length. Let $C(x)$ be the code
length to encode the sequence $x$ with a universal source coder.
Let $x$ be a training string and $y$ a test sequence. We can then
compare $\frac{C(x)}{|x|}$ with $\frac{C(y)}{|y|}$ (which could
be seen as a measure of entropy rate) or compare $C(xy)$ with $C(x)$
to detect change. The issue with this is that there are many completely
dissimilar sources that have the same entropy rate. As an example,
let the data be binary iid with the original source having $P(X=1)=\frac{1}{3}$
and the new source $P(X=1)=\frac{2}{3}$. Then the optimum code for
the original source and the optimum code for the new source have the
same length. On the other hand, atypicality will immediately distinguish
such sequences. 

\section{Binary IID Case}

\label{binaryiid.sec}

In order to clarify ideas, at first we consider a very simple model.
The typical model is iid binary with $P(X_{n}=1)=p$. The alternative
model class also binary iid but with $P(X_{n}=1)=\theta$, where $\theta$
is unknown. We want to decide if a given sequence $x^{l}$ is typical
or atypical. This can be stated as the hypothesis test problem
\begin{align*}
H_{0}: & \theta=p\\
H_{1}: & \theta\neq p
\end{align*}
This problem does not have an UMP (universal most powerful) test.
However, a common approach to solving this type of problem is the
GLRT (generalized likelihood ratio test) \cite{KayBook}. Let
\begin{align*}
P(b) & =P(X_{n}=b)\\
\hat{P}(b) & =\frac{N(b|x^{l})}{l}
\end{align*}
where $l$ is the sequence length and $N(b|x^{l})$ is the number
of $x_{n}=b\in\{0,1\}$. The GLRT is
\begin{align}
L & =\log\frac{\prod_{b=0}^{1}\hat{P}(b)^{N(b|x^{l})}}{\prod_{b=0}^{1}P(b)^{N(b|x^{l})}}\nonumber \\
 & =\sum_{b=0}^{1}N(b|x^{l})\log\frac{1}{l}N(b|x^{l})-\sum_{b=0}^{1}N(b|x^{l})\log P(b)\nonumber \\
 & =l\sum_{b=0}^{1}\hat{P}(b)\log\frac{1}{l}N(b|x^{l})-l\sum_{b=0}^{1}\hat{P}(b)\log P(b)\nonumber \\
 & =lD(\hat{p}\|p)\nonumber \\
\phi(x^{l}) & =\begin{cases}
1 & L>t\\
0 & L\leq t
\end{cases}\label{atyphyp.eq}
\end{align}
Where $D(\hat{p}\|p)=\sum_{b=0}^{1}\hat{P}(b)\log\frac{\hat{P}(b)}{P(b)}$
is the relative entropy \cite{CoverBook} and $t$ some threshold.
While the GLRT is a heuristic principle, it satisfies some optimality
properties, and in this case it is equal to the invariant UMP test
\cite{ScharfBook}, which can be considered an optimum solution under
certain constraints. Thus, it is reasonably to take this as the optimum
solution for this problem, and we do not need to appeal to Kolmogorov
or information theory to solve the problem.

The complications start if we consider sequences of variable length
$l$. The test (\ref{atyphyp.eq}) depends on the sequence length.
We need to choose a threshold $t(l)$ as a function of $l$, which
will then result in a false alarm probability $P_{FA}(t(l))$ and
detection probability $P_{D}(t(l))$. There is no obvious argument
for how to choose $t(l)$ from a hypothesis testing point of view;
we could choose $t$ independent of $l$, but that is just another
arbitrary choice.

We will consider this problem in the context of Definition \ref{atypdef.thm}.
In order to do so, we need to model the problem from a coding point
of view. We assume we have an (infinite) sequence of sequences of
variable length $l_{i}$, and these need to be encoded. We need to
encode each bit, and also to encode whenever a new sequence starts.
For typical encoding of the bits we can use a Shannon code, Huffman
code, arithmetic coding etc. The code length for a sequence of length
$l$ is 
\begin{align}
L_{t} & =N(1|x^{l})\log\frac{1}{p}+N(0|x^{l})\log\frac{1}{1-p}\nonumber \\
 & =l\left(\hat{p}\log\frac{1}{p}+(1-\hat{p})\log\frac{1}{1-p}\right)\label{TypicalCL.eq}
\end{align}
except for a small constant factor; here $\hat{p}=\hat{P}(1)=\frac{1}{l}\sum x_{i}$.
We also need to encode where a sequence ends and a new one starts.
For simplicity let us for now assume lengths are geometrically distributed.
We can then model the problem as one with three source symbols '0',
'1' and ',' with an iid distribution with $P(',')=\epsilon$, $P('0')=p-\frac{\epsilon}{2}$,
$P('1')=(1-p)-\frac{\epsilon}{2}$. If we assume $\epsilon$ is small,
the expression (\ref{TypicalCL.eq}) is still valid for the content
part, and to each sequence is added a constant $-\log\epsilon$ to
encode separators. To decide if a sequence is atypical according to
Definition \ref{atypdef.thm}, we can use the universal source coder
from \cite{CoverBook}: the source encodes first the number of ones
$k$; then it enumerates the sequences with $k$ ones, and transmits
the index of the given sequence. For analysis it is important to have
a simple expression for the code length. We can therefore use $L_{a}=lH(\hat{p})+\frac{1}{2}\log l$.
This is an approximation which is good for reasonably large $l$ and
it also reaches the lower bound in \cite{Rissanen86b,Shamir06}. The
source-coder also needs to inform the decoder that the following is
an atypical sequence (so that it knows to use the atypical decoder
rather than the typical encoder), and where it ends. For the former
we can use a '.' to indicate the start of an atypical sequence rather
than the ',' for typical sequences. If the probability that a sequence
is atypical is $\delta\ll1$, $P('.')=\delta\epsilon$ and $P(',')=(1-\delta)\epsilon\approx\epsilon$.
The code length for a '.' now is $-\log\epsilon-\log\delta$. To mark
the end of the atypical sequence we could again insert a '.' or a
','. But the code for either is based on the distribution of lengths
of \emph{typical} sequences, which we assume known, whereas we would
have no knowledge of the length of atypical sequences. Instead it
seems more reasonable to encode the length of the specific atypical
sequence. As argued in \cite{Rissanen83,Elias75} this can be done
with $\log^{*}l+\log c$, where $c$ is a constant and 
\begin{equation}
\log^{*}(l)=\log l+\log\log l+\log\log\log l+\cdots
\end{equation}
where the sum continues as long as the argument to the log is positive.
To summarize we have
\begin{align}
L_{t} & =l\left(\hat{p}\log\frac{1}{p}+(1-\hat{p})\log\frac{1}{1-p}\right)-\log\epsilon\nonumber \\
L_{a} & =lH(\hat{p})+\frac{1}{2}\log l+\log^{*}l+\log c-\log\epsilon-\log\delta\nonumber \\
 & \approx lH(\hat{p})+\frac{3}{2}\log l-\log\epsilon+\tau\nonumber \\
\tau & =-\log\delta+\log c\label{LtLaiid.eq}
\end{align}
The criterion for a sequence to be atypical is $L_{a}<L_{t}$, which
easily seen to be equivalent to
\begin{equation}
D(\hat{p}\|p)>\frac{\tau+\frac{3}{2}\log l}{l}\label{Diid.eq}
\end{equation}
If the lengths are fixed, this reduces to (\ref{atyphyp.eq}). But
if the lengths are variable, (\ref{Diid.eq}) provides a threshold
as a function of $l$. The term $\frac{3}{2}\log l$ ensures that
$\lim_{l\to\infty}P_{FA}(l)=0$, which seems reasonable. If instead
$D(\hat{p}\|p)>\frac{\tau}{l}$ is used, it is easy to see that $\lim_{l\to\infty}P_{FA}(l)>0$.
Except for this property, the term $\frac{3}{2}\log l$ might seem
arbitrary, e.g., why $\frac{3}{2}$? But it is based on solid theory,
and as will be seen later it has several important theoretical properties.

\begin{figure}[tbh]
\includegraphics[width=3.5in]{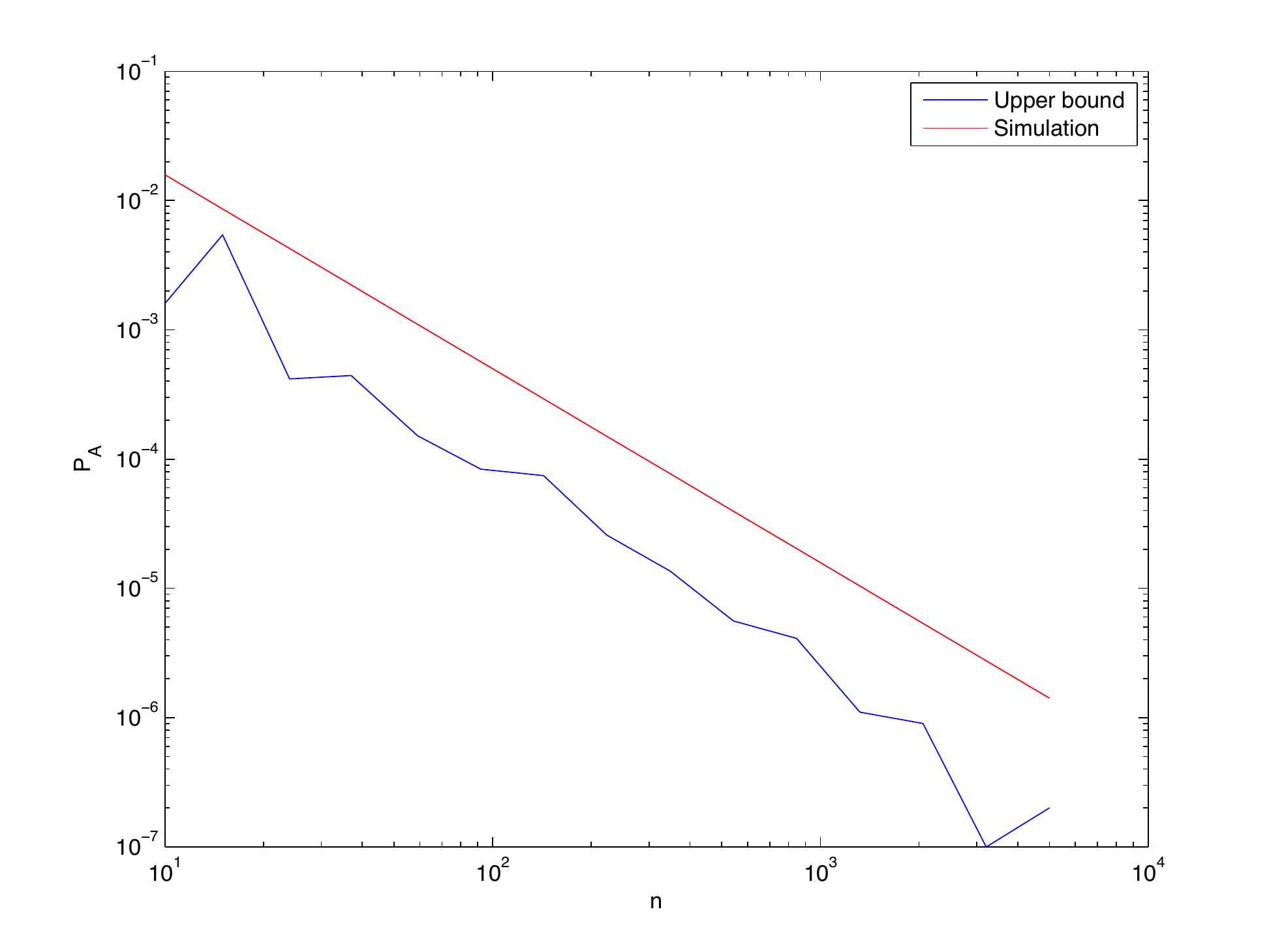}

\caption{\label{PA.fig}Simulated $P_{A}$ and the Upper bound for $\tau=1,p=0.3$.}
\end{figure}

We will examine the criterion (\ref{Diid.eq}) in more detail. The
inequality (\ref{Diid.eq}) gives two thresholds for $\hat{p}$, 
\begin{align*}
\hat{p} & >p_{+}\\
\hat{p} & <p_{-}
\end{align*}
Where $0<p_{-}<p<p_{+}<1$. It is impossible to find explicit expressions
for $p_{\pm}$, but it is clear that 
\begin{align*}
p_{\pm}\to p & \quad\mbox{ as }l\to\infty.
\end{align*}
Therefore, for $l$ large, we can replace $D(\hat{p}\|p)$ with a
series expansion. We then end up with the more explicit criterion
\begin{align}
\frac{(p-\hat{p})^{2}}{pq\ln4} & >\frac{1}{l}(\tau+\frac{3}{2}\log l)\nonumber \\
\left|\hat{p}-p\right| & >\Delta\tau\stackrel{\cdot}{=}\sqrt{\frac{pq\ln4}{l}}\sqrt{\tau+\frac{3}{2}\log l}\label{approxcrit.eq}
\end{align}
In the following we will use this as it is considerably simpler to
analyze. We can also write this as 
\begin{align}
\left|\frac{\sum_{i=1}^{l}x_{i}-p}{\sqrt{pql}}\right| & >\sqrt{2\tau\ln2+3\ln l}\label{CLT.eq}
\end{align}
Now, if not for the term $3\ln l$, this would be a central limit
type of statement, and the probability that a sequence is classified
as (intrinsically) atypical would be 
\begin{align}
P_{A} & \approx2Q\left(\sqrt{2\ln2\tau}\right)\label{CLT2.eq}
\end{align}
independent of $l$. Our main interest is exactly the the dependency
on $l$, which is given by the following Theorem 
\begin{thm}
\label{1storderupper.thm}Consider an iid $\{0,1\}$-sequence. Let
$P_{A}(l)$ be the probability that a sequence of length $l$ is classified
as \emph{intrinsically} atypical according to (\ref{approxcrit.eq}).
Then $P_{A}(l)$ is bounded by
\begin{align}
P_{A}(l)\leq & 2^{-\tau+1}\frac{1}{l^{3/2}}K(l,\tau)\label{PAupper.eq}\\
\forall\tau:\lim_{l\to\infty}K(l,\tau)= & 1\nonumber 
\end{align}
For $p=\frac{1}{2}$ this can be strengthened to
\begin{align}
P_{A}(l)\leq & 2^{-\tau+1}\frac{1}{l^{3/2}}\label{PAupperunf.eq}
\end{align}
These bounds are tight in the sense that
\begin{equation}
\lim_{l\to\infty}\frac{\ln P_{A}(l)}{-\frac{3}{2}\ln l}=1\label{PAlower.eq}
\end{equation}
\end{thm}
\begin{IEEEproof}
The Chernoff bound (e.g., \cite{PoorBook}) states
\begin{align*}
P_{A}(l)=P\left(\left|\hat{p}-p\right|>\Delta\tau\right)= & 2P\left(\sum_{i=1}^{l}X_{i}\geq lp+b\right)\\
\leq & 2\inf_{s>0}\left\{ e^{-lsp-sb}M_{X}(s)^{l}\right\} 
\end{align*}
Where (as usual, $q=1-p$)
\begin{align*}
b= & \sqrt{\vphantom{\tau+\frac{3}{2}\log l}lpq\ln4}\sqrt{\tau+\frac{3}{2}\log l}
\end{align*}
and $M_{S}(s)$ is the moment generating function of $X_{i}$, which
for a Bernoulli random variable is
\begin{align*}
M_{X}(s)= & pe^{s}+q
\end{align*}
Then

\begin{align*}
\frac{1}{2}P_{A}(l)\leq & \inf_{s>0}\left\{ \exp\left(-s\left(pl+b\right)\right)\left(pe^{s}+q\right)^{l}\right\} 
\end{align*}
Minimizing over $s$ gives
\begin{align*}
\frac{1}{2}P_{A}(l)\leq & \left(\frac{lq}{lq-b}\right)^{l}\left(\frac{q(lp+b)}{p(lq-b)}\right)^{-lp-b}
\end{align*}
or 
\begin{align}
\ln\frac{1}{2}P_{A}(l)\leq & l\ln\left(\frac{lq}{lq-b}\right)+\left(-lp-b\right)\ln\left(\frac{q(lp+b)}{p(lq-b)}\right)\nonumber \\
= & l\ln\left(1+\frac{b}{lq-b}\right)+\left(-lp-b\right)\ln\left(1+\frac{b}{p(lq-b)}\right)\nonumber \\
\leq & \frac{b^{2}\left(3l^{2}q^{2}p+lb(7p^{2}-6p-3)+b^{2}(6p+3)\right)}{6p^{2}(b-lq)^{3}}\nonumber \\
\leq & -\frac{b^{2}}{2lpq}+O\left(1\right)\frac{b^{3}}{l^{2}}\nonumber \\
= & -\tau\ln2-\frac{3}{2}\ln l+O\left(1\right)\frac{\ln^{3/2}l}{\sqrt{l}}\tau^{3/2},\label{PAproofbound.eq}
\end{align}
where we have used $x-\frac{x^{2}}{2}\leq\ln(1+x)\leq x-\frac{x^{2}}{2}+\frac{x^{3}}{3}$
for $x\geq0$. The equation (\ref{PAproofbound.eq}) directly leads
to (\ref{PAupper.eq}). 

For $p=\frac{1}{2}$ Hoeffding's inequality \cite{hoeffding1963probability}
gives the bound
\begin{align}
P_{A}(l) & \leq2\exp\left(-2\frac{b^{2}}{l}\right)\nonumber \\
 & =2\exp\left(-\ln2\left(\tau+\frac{3}{2}\log l\right)\right)\label{Hoeffding.eq}
\end{align}
for $p=\frac{1}{2}$ this is tighter than (\ref{PAproofbound.eq}). 

For the lower bound we use moderate deviations from \cite{DemboBook}.
Define $\tilde{X}_{i}=\frac{X_{i}-p}{\sqrt{pq}}$. We can then rewrite
(\ref{CLT.eq}) as
\begin{align*}
\left|\frac{\sum_{i=1}^{l}\tilde{X}_{i}}{\sqrt{l(2\tau\ln2+3\ln l)}}\right| & >1
\end{align*}
We define $a_{l}=\frac{1}{2\tau\ln2+3\ln l}$, which satisfies $\lim_{l\to\infty}a_{l}=0$,
$\lim_{l\to\infty}la_{l}=\infty$. Using this as $a_{l}$ in \cite[Theorem 3.7.1]{DemboBook}
gives 
\begin{align*}
\lefteqn{\liminf_{l\to\infty}\frac{1}{2\tau\ln2+3\ln l}\ln P\left(\left|\frac{\sum_{i=1}^{l}\tilde{X}_{i}}{\sqrt{l(2\tau\ln2+3\ln l)}}\right|>1\right)}\\
 & =\liminf_{l\to\infty}\frac{1}{3\ln l}\ln P\left(\left|\frac{\sum_{i=1}^{l}\tilde{X}_{i}}{\sqrt{l(2\tau\ln2+3\ln l)}}\right|>1\right)\\
 & \geq-\frac{1}{2}
\end{align*}
Together with the upper bound, this gives (\ref{PAlower.eq}).

\end{IEEEproof}
Figure \ref{PA.fig} compares the upper bound with simulations. 

We can also bound the miss probability for \emph{extrinsically} atypical
sequences as follows
\begin{thm}
\label{iiddetection.thm}Suppose that the typical sequence is iid
$\{0,1\}$-sequence with $P(X_{n}=1)=p$. Let the test sequence by
iid with $P(X_{n}=1)=p_{a}$. The probability that the test sequence
is missed according to criterion (\ref{approxcrit.eq}) is upper bounded
by 
\begin{align}
P_{M}(l)\leq & 2^{-\tau}\frac{1}{l^{3/2}}\left(\frac{q_{a}p}{p_{a}q}\right)^{\sqrt{lpq(2\tau\ln2+3\ln l)}}\nonumber \\
 & \times\left(\frac{q_{a}^{p-1}q^{p+1}}{p_{a}^{p}p^{p}}\right)^{-l}K(l,\tau)\\
\forall\tau:\lim_{l\to\infty}K(l,\tau)= & 1\nonumber 
\end{align}
\end{thm}
\begin{IEEEproof}
We may assume that $p_{a}<p$. Similarly to the proof of Theorem \ref{1storderupper.thm}
the Chernoff bound is

\begin{align*}
P_{M}(l)\leq & \inf_{s>0}\left\{ \exp\left(-s\left(pl-b\right)\right)\left(p_{a}e^{s}+q_{a}\right)^{l}\right\} 
\end{align*}
Minimizing over $s$ gives
\begin{align*}
P_{M}(l)\leq & \left(\frac{lq_{a}}{lq+b}\right)^{l}\left(\frac{q_{a}(lp-b)}{p_{a}(lq+b)}\right)^{-lp+b}
\end{align*}
or
\begin{align*}
\ln P_{M}(l)\leq & l\ln\left(\frac{lq_{a}}{lq+b}\right)+\left(-lp+b\right)\ln\left(\frac{q_{a}(lp-b)}{p_{a}(lq+b)}\right)\\
\leq & l\ln\left(\frac{q_{a}}{q}\right)-lp\ln\left(\frac{q_{a}}{p_{a}}\right)-lp\ln\left(\frac{q}{p}\right)\\
 & +b\left(\ln\left(\frac{p}{q}\right)+\ln\left(\frac{q_{a}}{p_{a}}\right)\right)-\frac{b^{2}}{2lpq}+O\left(\frac{b^{3}}{l^{2}}\right)
\end{align*}
using series expansions.
\end{IEEEproof}

\subsection{Hypothesis testing interpretation}

The solution (\ref{Diid.eq}) may seem arbitrary, but it has a nice
interpretation in terms of hypothesis testing \cite{LehmannBook}.
Return to the solution (\ref{atyphyp.eq}). That solution gives a
test for a given $l$. However, the problem is that it does not reconcile
tests for different $l$. One way to solve that issue is to consider
$l$ a random variable, i.e., introducing a prior distribution in
the Bayesian sense. Let the prior distribution of $l$ be $P_{L}(l)$.
The equation (\ref{atyphyp.eq}) now becomes
\begin{align*}
L & =\log\frac{\prod_{b=0}^{1}\hat{P}(b)^{N(b|x^{l})}P_{L}(l)}{\prod_{b=0}^{1}P(b)^{N(b|x^{l})}P_{L}(0)}\\
 & =l\sum_{b=0}^{1}\hat{P}(b)\log\frac{1}{l}N(b|x^{l})\\
 & -l\sum_{b=0}^{1}\hat{P}(b)\log P(b)+\log P_{L}(l)-\log P_{L}(0)\\
 & =lD(\hat{p}\|p)+\log P_{L}(l)-\log P_{L}(0)
\end{align*}
The hypothesis test now is
\begin{align}
D(\hat{p}\|p)> & \frac{\tau+\log P_{L}(0)-\log P_{L}(l)}{l}\label{hyp1.eq}
\end{align}
Of course, the problem is that we don't know $P(l)$. Still, compare
that with (\ref{Diid.eq}) without the approximations,
\begin{align}
D(\hat{p}\|p)> & \frac{\tau+\frac{1}{2}\log l+c+\log^{*}l}{l}\label{hyp2.eq}
\end{align}
To the term $c+\log^{*}l$ corresponds a distribution on the integers,
namely $Q(l)$ in \cite[(3.6)]{Rissanen83}. Except for the term $\frac{1}{2}\log l$,
the equations (\ref{hyp1.eq}) and (\ref{hyp2.eq}) are identical
if we use the prior distribution $P_{L}(l)=Q(l)$. Rissanen \cite{Rissanen83}
argues that the distribution $Q(l)$ is the most reasonable distribution
on the integers when we have really no prior knowledge, mainly from
a coding point of view. This therefore seems a reasonable distribution
for $P(l)$. What about the term $\frac{1}{2}\log l$? The model for
the non-null hypothesis has one unknown parameter, $p$, so that it
is more complex than the null hypothesis. We have to account for this
additional complexity. Our goal is to find an explanation for atypical
sequences among a large class of explanations, not just the distribution
of zeros and ones. If there is no ``penalty'' for finding a complex
explanation, any data can be explained, and all data will by atypical.
This is Occam's razor \cite{CoverBook}. The ``penalty'' for one
unknown parameter as argued by Rissanen is exactly $\frac{1}{2}\log l$.
We therefore have the following explanation for (\ref{Diid.eq}),
\begin{fact}
The criterion (\ref{Diid.eq}) can be understood as a hypothesis test
with prior distribution $Q(l)$ \cite{Rissanen83} and penalty $\frac{1}{2}\log l$
for the unknown parameter.
\end{fact}
Seen in this light, Theorem \ref{1storderupper.thm} is not surprising.
In (\ref{Diid.eq}) we have replaced $\frac{1}{2}\log l+\log^{*}l$
with $\frac{3}{2}\log l$, which implicitly corresponds to the prior
distribution $P_{L}(l)\sim l^{-3/2}$, which is exactly the distribution
seen in (\ref{PAupper.eq}).

\subsection{\label{Subsequences.sec}Atypical subsequences}

One problem where we believe our approach excels is in finding atypical
\emph{subsequences }of long sequences. The difficulty in find atypical
subsequences is that we may have short subsequences that deviate much
from the typical model, and long subsequences that deviate little.
How do we choose among these? Definition \ref{atypdef.thm} gives
a precise answer. For the formal problem statement, consider a sequence
$\{x_{n},n=-\infty,\ldots,\infty\}$ from a finite alphabet $\mathcal{A}$
(where in this section $\mathcal{A}=\{0,1\}$). The sequence is generated
according to a probability law $\mathcal{P}$, which is known. In
this sequence is embedded (infrequent) finite subsequences $\mathcal{X}_{i}=\{x_{n},n=n_{i},\dots,n_{i}+l_{i}-1\}$
from the finite alphabet $\mathcal{A}$, which are generated by an
alternative probability law $\tilde{\mathcal{P}}_{\theta}$. The probability
law $\tilde{\mathcal{P}}_{\theta}$ is unknown, but it might be known
to be from a certain class of probability distributions, for example
parametrized by the parameter $\theta$. Each subsequence $\mathcal{X}_{i}$
may be drawn from a different probability law. The problem we consider
is to isolate these subsequences, which we call atypical subsequences.
In this section, as above, we will assume both $\mathcal{P}$ and
$\tilde{P}_{\theta}$ are binary iid.

The solution is very similar to the one for variable length sequences
above. The atypical subsequences are encoded with the universal source
coder from \cite{CoverBook} with a code length $L_{a}=lH(\hat{p})+\frac{1}{2}\log l$.
The start of the sequence is encoded with an extra symbol '.' which
has a code length $-\log P('.')$ and the length is encoded in $\log^{*}l$
bits. In conclusion we end up with exactly the same criterion as (\ref{Diid.eq}),
repeated here
\begin{equation}
D(\hat{p}\|p)>\frac{\tau+\frac{3}{2}\log l}{l}\label{Diid.eq-2}
\end{equation}
The only difference is that $\tau$ has a slight different meaning.

For the subsequence problem, a central question is what the probability
is that a given sample $x_{n}$ is part of an (intrinsically) atypical
subsequence. Notice that there are infinitely many subsequences that
can contain $x_{n}$, and each of these have a probability of being
atypical given by Theorem \ref{1storderupper.thm}. 

We can obtain an upper bound as follows. Let us say that $X_{n}$
has been determined to be part of an atypical sequence $\mathcal{X}_{i}$.
It is clear that the sequence $\mathcal{X}_{i}$ must also be atypical
according to (\ref{Diid.eq-2}). Therefore, we can upper bound the
probability $P_{A}(X_{n})$ that $X_{n}$ is part of an atypical sequence
with the probability of the event (\ref{Diid.eq-2}), using the approximate
criterion (\ref{approxcrit.eq}), 
\begin{align*}
\exists n_{1}\leq n<n_{1}+l:\left|\frac{\sum_{i=n_{1}}^{n_{1}+l-1}X_{i}-p}{\sqrt{pql}}\right| & >\sqrt{2\tau\ln2+3\ln l}
\end{align*}
We can rewrite this as
\begin{align*}
\exists n_{1}\leq n<n_{1}+l:\left|\frac{\sum_{i=n_{1}}^{n_{1}+l-1}X_{i}-p}{\sqrt{pql}}\right| & >\sqrt{2\tau\ln2+3\ln l}\\
\exists n_{1}\leq n<n_{1}+l:\left|\sum_{i=n_{1}}^{n_{1}+l-1}X_{i}-p\right| & >\sqrt{lpq\ln2(2\tau+3\log l)}
\end{align*}
We could upper bound this with a union bound using Theorem \ref{1storderupper.thm}.
However, it is quickly seen that this does not converge. The problem
is that the events in the union bound are highly dependent, so we
need a slightly more refined approach; this results in the following
Theorem
\begin{thm}
\label{Upper.thm}Consider the case $p=\frac{1}{2}$. The probability
$P_{A}(X_{n})$ that a given sample $X_{n}$ is part of an atypical
subsequence is upper bounded by
\begin{equation}
P_{A}(X_{n})\leq(K_{1}\sqrt{\tau}+K_{2})2^{-\tau}\label{PAsampleup.eq}
\end{equation}
for some constants $K_{1},K_{2}$. 
\end{thm}
\begin{IEEEproof}
Without loss of generality we can assume $n=0$. For some $l_{0}>0$
let $\mathcal{I}_{l_{0}}$ be the set of subsequences containing $X_{0}$
of length $l\leq l_{0}$. For $i\in\mathcal{I}_{l_{0}}$ let $l(i)$
be the length of the subinterval. From Theorem \ref{1storderupper.thm}
we know that $P_{A}(i)\leq2^{-\tau+1}\frac{1}{l^{3/2}}K(l,\tau)$
and therefore
\[
\sum_{i\in\mathcal{I}_{l_{0}}}P_{A}(i)\leq K2^{-\tau}
\]
for some constant $K$. This argument does not work if we allow arbitrarily
long subsequences, because the sum is divergent. However, we can write
\[
P_{A}(X_{0})\leq\sum_{i\in\mathcal{I}_{l_{0}}}P_{A}(i)\leq K2^{-\tau}+P_{A,l_{0}}(X_{0})
\]
where $P_{A,l_{0}}(X_{0})$ is the probability that $X_{0}$ is in
an atypical subsequence of at least length $l_{0}$. The proof will
be to bound $P_{A,l_{0}}(X_{0})$.

Define the following events
\begin{align*}
\overline{A}(n_{1},l)= & \left\{ \sum_{i=n_{1}}^{n_{1}+l-1}X_{i}-p>\sqrt{lpq\ln2(2\tau+3\log l)}\right\} \\
\underline{A}(n_{1},l)= & \left\{ \sum_{i=n_{1}}^{n_{1}+l-1}X_{i}-p<-\sqrt{lpq\ln2(2\tau+3\log l)}\right\} 
\end{align*}
For $p=\frac{1}{2}$ we can rewrite
\begin{align}
\lefteqn{\sum_{i=n_{1}}^{n_{1}+l-1}X_{i}-p>\sqrt{lpq\ln2(2\tau+3\log l)}}\nonumber \\
= & \sum_{i=n_{1}}^{n_{1}+l-1}(2X_{i}-1)>\sqrt{l\ln2(2\tau+3\log l)}\label{RW.eq}
\end{align}
For ease of notation define
\begin{align*}
\upsilon(l)= & \left\lceil \sqrt{l\ln2(2\tau+3\log l)}\right\rceil 
\end{align*}
Then using the union bound we can write
\begin{align*}
P_{A}(X_{0})\leq & \sum_{n_{1}=-\infty}^{-1}P\left(\bigcup_{l=-n_{1}+1}^{\infty}\overline{A}(n_{1},l)\right)\\
 & +\sum_{n_{1}=-\infty}^{-1}P\left(\bigcup_{l=-n_{1}+1}^{\infty}\underline{A}(n_{1},l)\right)\\
\leq & 2\sum_{n_{1}=-\infty}^{-1}P\left(\bigcup_{l=-n_{1}+1}^{\infty}\overline{A}(n_{1},l)\right)\\
= & 2\sum_{n_{1}=-\infty}^{-1}1-P\left(\bigcap_{l=-n_{1}+1}^{\infty}\overline{A}^{c}(n_{1},l)\right)\\
= & 2\sum_{n_{1}=-\infty}^{-1}\left(1-\vphantom{\prod_{l=n-n_{1}}^{\infty}P\left(\overline{A}^{c}(n_{1},l)\left|\bigcap_{\ell=n-n_{1}}^{l-1}\overline{A}^{c}(n_{1},\ell)\right.\right)}\right.\\
 & \left.\prod_{l=-n_{1}+1}^{\infty}P\left(\overline{A}^{c}(n_{1},l)\left|\bigcap_{\ell=n-n_{1}+1}^{l-1}\overline{A}^{c}(n_{1},\ell)\right.\right)\right)
\end{align*}
where we have excluded the length one sequence consisting of $X_{0}$
itself. Now consider $P\left(\overline{A}^{c}(n_{1},l)\left|\bigcap_{\ell=-n_{1}+1}^{l-1}\overline{A}^{c}(n_{1},\ell)\right.\right)=1-P\left(\overline{A}(n_{1},l)\left|\bigcap_{\ell=-n_{1}+1}^{l-1}\overline{A}^{c}(n_{1},\ell)\right.\right)$. 

We can think of $S_{l}=\sum_{i=n_{1}}^{n_{1}+l-1}(2X_{i}-1)$ as a
simple random \cite{GrimmettBook} , and we will use this to upper
bound the probability $P\left(\overline{A}(n_{1},l)\left|\bigcap_{\ell=-n_{1}+1}^{l-1}\overline{A}^{c}(n_{1},\ell)\right.\right)$.
This probability can be interpreted as the probability that the random
walk passes $\upsilon(l)$ given that it was below $\upsilon(\ell)$
at times $-n_{1}<\ell<l$. But since the random walk can increase
by at most one, and since the threshold is increasing with $l$, that
means that at time $l$ we must have $S_{l}=\upsilon(l)$. Furthermore,
it is easy to see that the probability is upper bounded by the probability
that $S_{l}=\upsilon(l)$ given that the random walk is below $\upsilon(l)$
at times $-n_{1}<\ell<l$. Thus
\begin{align}
\lefteqn{P\left(\overline{A}(n_{1},l)\left|\bigcap_{\ell=-n_{1}+1}^{l-1}\overline{A}^{c}(n_{1},\ell)\right.\right)}\nonumber \\
 & \leq P\left(S_{l}=\upsilon(l)\left|S_{\ell}<\upsilon(l),-n_{1}<\ell<l\right.\right)\nonumber \\
 & =\frac{P\left(S_{l}=\upsilon(l),S_{\ell}<\upsilon(l),-n_{1}<\ell<l\right)}{P\left(S_{\ell}<\upsilon(l),-n_{1}<\ell<l\right)}\nonumber \\
 & \leq\frac{P\left(S_{l}=\upsilon(l),S_{\ell}<\upsilon(l),-n_{1}<\ell<l\right)}{P\left(S_{\ell}<\upsilon(l),0\leq\ell<l\right)}\label{PA1.eq}
\end{align}
The denominator can be interpreted as the probability that the maximum
of the random walk stays below $\upsilon(l)$, which by Theorem \ref{1storderupper.thm}
can be expressed by
\begin{align*}
P_{D}(l) & =P\left(S_{\ell}<\upsilon(l),0\leq\ell<l\right)\\
 & =1-2P(S_{l-1}\geq\upsilon(l+1)-1)-P(S_{l-1}=\upsilon(l)-1)\\
 & \geq1-2^{-\tau+c}l^{-3/2}\\
 & \geq\frac{1}{2}
\end{align*}
for $\tau$ and $l$ sufficiently large, and where $c$ is some constant.
Since, as discussed at the start of the proof, we can assume that
$l\geq l_{0}$, we can choose $l_{0}$ large enough that this is satisfied;
furthermore, since $P_{N}(l)$ is increasing in $\tau$, we can choose
$l_{0}$ independent of $\tau$ as long as $\tau$ is sufficiently
large.

We will next upper bound the numerator in (\ref{PA1.eq}). This is
the probability that we have a path that has stayed below $\upsilon(l)$
at steps $-n_{1}<\ell<l$, but then at step $l$ hits $\upsilon(l)$.
We will count such paths. We divide them into two groups that we count
separately. The first group are all paths that start at zero and hit
$\upsilon(l)$ \emph{first time} after $l$ steps. The second group
is more easily described in reverse time. Those are paths that start
at $\upsilon(l)$ at step $l$, then stay below $\upsilon(l)$ until
time $\tilde{n}<0$, when they hit $\upsilon(l)$ again, and finally
hit 0 at time $n_{1}$. According to \cite[Section 3.10]{GrimmettBook}
we can count all these paths by

\begin{equation}
N=\frac{\upsilon(l)}{l}N_{l}(0,\upsilon(l))+\sum_{t=\upsilon(l)}^{-n_{1}+1}\frac{1}{l-t-1}N_{l-t-1}(1,0)N_{t}(0,\upsilon(l))\label{Npaths.eq}
\end{equation}
Where $N_{n}(a,b)$ are the number of length $n$ paths between $a$
and \textbf{$b$. }

We need to upper bound the probability $P(S_{n}=k)$ that a path starting
a $0$ hits $k$ after $n$ steps. We use \cite[Section 3.10]{GrimmettBook}
and \cite[13.2]{CoverBook} to get
\begin{align*}
P\left(S_{n}=k\right)= & N_{n}(0,k)2^{-n}\\
= & \left(\begin{array}{c}
n\\
\frac{1}{2}(n+k)
\end{array}\right)2^{-n}\\
\leq & \sqrt{\frac{n}{\pi\frac{1}{4}(n+k)(n-k)}}2^{nH\left(\frac{\frac{1}{2}(n+k)}{n}\right)}2^{-n}\\
= & \sqrt{\frac{4n}{\pi(n^{2}-k^{2})}}2^{nH\left(\frac{\frac{1}{2}(n+k)}{n}\right)}2^{-n}
\end{align*}
We can bound the power of the exponent to 2 as follows
\begin{align}
\lefteqn{nH\left(\frac{\frac{1}{2}(n+k)}{n}\right)-n}\nonumber \\
= & n\left(H\left(\frac{1}{2}+\frac{1}{2}\frac{k}{n}\right)-1\right)\nonumber \\
\leq & -\frac{2}{\ln2}n\left(\frac{1}{2}\frac{k}{n}\right)^{2}\nonumber \\
= & -\frac{1}{2\ln2}\frac{k^{2}}{n}\label{Hbound.eq}
\end{align}
Thus,
\begin{align*}
P\left(S_{n}=k\right)\leq & \frac{2}{\sqrt{\pi n}}e_{2}\left(-\frac{1}{2\ln2}\frac{k^{2}}{n}\right)
\end{align*}
where $e_{2}(x)=2^{x}$.

We will use this to bound the probability of set of paths in the second
term in (\ref{Npaths.eq}). We can bound

\begin{align*}
P_{2}(n_{1},l) & =\sum_{t=\upsilon(l)}^{-n_{1}+1}\frac{1}{l-t-1}N_{l-t-1}(1,0)N_{t}(0,\upsilon(l))2^{-(l-1)}\\
 & \leq\sum_{t=\upsilon(l)}^{-n_{1}+1}\frac{1}{l-t-1}\frac{2}{\sqrt{\pi(l-t-1)}}\\
 & \quad\times e_{2}\left(-\frac{1}{2\ln2}\frac{1}{l-t-1}\right)N_{t}(0,\upsilon(l))2^{-t}\\
 & \leq\frac{4}{\pi(l+n_{1}-2)^{3/2}}\sum_{t=\upsilon(l)}^{-n_{1}+1}P(S_{t}=\upsilon(l))
\end{align*}
Here the sum $\sum_{t=\upsilon(l)}^{-n_{1}+1}P(S_{t}=\upsilon(l))$
when looked at in reverse time can be interpreted as the probability
of a path starting at $\upsilon(l)$ hits zero before time $-n_{1}+1$.
We can the write this as (See \cite[Section 3.10]{GrimmettBook})
\begin{align*}
\sum_{t=\upsilon(l)}^{-n_{1}+1}P(S_{t}=\upsilon(l)) & =P(M_{-n_{1}+1}\geq\upsilon(l))\\
 & \leq2P(S_{-n_{1}+1}\geq\upsilon(l))
\end{align*}
We can use the proof of Theorem \ref{1storderupper.thm}, specifically
(\ref{Hoeffding.eq}) to bound this by 
\[
P(M_{-n_{1}+1}\geq\upsilon(l))\leq\exp\left(-\frac{2\upsilon(l)^{2}}{-n_{1}+1}\right)
\]
Then
\begin{align*}
P_{2}(n_{1},l) & \leq\frac{K}{(l+n_{1}-2)^{3/2}}\exp\left(-\frac{\upsilon(l)^{2}}{2(-n_{1}+1)}\right)\sqrt{-n_{1}+1}
\end{align*}

We will next bound the probability of the paths in the first term
in (\ref{Npaths.eq}). We have
\begin{align*}
P\left(S_{l}=\upsilon(l)\right)\leq & \sqrt{\frac{4l}{\pi(l^{2}-\upsilon(l)^{2})}}e_{2}\left(-\frac{1}{2\ln2}\frac{\upsilon(l)^{2}}{l}\right)\\
 & \sqrt{\frac{4}{\pi l\left(1-\frac{\upsilon(l)^{2}}{l^{2}}\right)}}e_{2}\left(-\frac{1}{2\ln2}\frac{\upsilon(l)^{2}}{l}\right)\\
= & \frac{2}{\sqrt{\pi\left(1-\frac{\upsilon(l)^{2}}{l^{2}}\right)}}2^{-\tau}l^{-2}
\end{align*}
and
\begin{align*}
P_{1}(l) & =\frac{\upsilon(l)}{l}P\left(S_{l}=\upsilon(l)\right)\\
\leq & \sqrt{\frac{\ln2(2\tau+3\log l)}{l}}\frac{2}{\sqrt{\pi\left(1-\frac{\upsilon(l)^{2}}{l^{2}}\right)}}2^{-\tau}l^{-2}\\
\leq & \sqrt{\frac{8\tau\ln2}{\pi\left(1-\frac{\upsilon(l)^{2}}{l^{2}}\right)}}2^{-\tau}l^{-5/2}\\
 & +\sqrt{\frac{12\ln l}{\pi\left(1-\frac{\upsilon(l)^{2}}{l^{2}}\right)}}\sqrt{\frac{4}{\pi}}2^{-\tau}l^{-5/2}
\end{align*}
Thus

\begin{align*}
\lefteqn{\ln\left(\prod_{l=-n_{1}+1}^{\infty}P\left(\overline{A}^{c}(n_{1},l)\left|\bigcap_{\ell=-n_{1}+1}^{l-1}\overline{A}^{c}(n_{1},\ell)\right.\right)\right)}\\
\geq & \sum_{l=-n_{1}+1}^{\infty}\ln\left(1-\frac{P_{1}(l)-P_{2}(n_{1},l)}{P_{D}(l)}\right)\\
\geq & K\sum_{l=-n_{1}+1}^{\infty}-P_{1}(l)-P_{2}(n_{1},l)\\
\doteq & S(-n_{1},\tau)
\end{align*}
and
\begin{align}
P_{A}(X_{0})\leq & 2\sum_{n_{1}=-\infty}^{-1}1-e^{S(-n_{1},\tau)}\nonumber \\
\leq & 2K\sum_{n_{1}=-\infty}^{-1}\sum_{l=-n_{1}+1}^{\infty}P_{1}(l)\nonumber \\
 & +2K\sum_{n_{1}=-\infty}^{-1}\sum_{l=-n_{1}+1}^{\infty}P_{2}(n_{1},l)\label{PASS.eq}
\end{align}
where $K>0$ is some constant. 

First we evaluate the sum of $P_{1}(l)$. The term $\frac{\upsilon(l)^{2}}{l^{2}}$
is decreasing in $l$, so for sufficiently large $l_{1}$, $\frac{\upsilon(l)^{2}}{l^{2}}\leq\frac{1}{2}$.
We can evaluate the sum separately for $l_{0}\leq l\leq l_{1}$ and
for $l>l_{1}$. Convergence depends only on the latter tail. The threshold
$l_{1}$ is increasing with $\tau$. If for example we put $l_{1}=8\tau\ln2$
, i.e., proportional to $\tau$, we have $\frac{\upsilon(l)^{2}}{l^{2}}\leq\frac{1}{2}$
for $\tau>10$. Therefore
\[
\sum_{l=l_{0}}^{l_{1}}P_{1}(l)\leq K\tau2^{-\tau}
\]
For $l>l_{1}$ we can write
\begin{align*}
P_{1}(l) & =\sqrt{\frac{16\tau\ln2}{\pi}}2^{-\tau}l^{-5/2}+\sqrt{\frac{24\ln l}{\pi}}\sqrt{\frac{4}{\pi}}2^{-\tau}l^{-5/2}
\end{align*}
Then (for $-n_{1}+1>l_{1}$ ) 
\begin{align}
\lefteqn{K\sum_{l=-n_{1}+1}^{\infty}P_{1}(l)}\nonumber \\
\leq & k_{1}2^{-\tau}\sqrt{\frac{\ln(n-n_{1})}{(-n_{1})^{3}}}\nonumber \\
 & +k_{2}2^{-\tau}\mathrm{erfc}\left(\sqrt{\frac{3}{2}\ln(-n_{1})}\right)\nonumber \\
 & +k_{3}2^{-\tau}\sqrt{\frac{\tau}{(-n_{1})^{3}}}\label{PAupn.eq}
\end{align}
where $k_{i}>0$ are some constants and where we have used
\begin{align*}
\sum_{l=k}^{\infty}l^{-5/2}\leq & \int_{k-1}^{\infty}x^{-5/2}dx=\frac{2}{3(k-1)^{3/2}}\\
\sum_{l=k}^{\infty}\sqrt{\ln l}l^{-5/2}\leq & \int_{k-1}^{\infty}\sqrt{\ln x}x^{-5/2}dx\\
= & \frac{1}{9}\left(\sqrt{6\pi}\text{erfc}\left[\sqrt{\frac{3}{2}\ln k-1}\right]+\frac{6\sqrt{\ln k-1}}{(k-1)^{3/2}}\right)
\end{align*}
as it can be verified that all three sums, when (\ref{PAupn.eq})
is inserted in (\ref{PASS.eq}), are convergent, using $\sum_{k=1}^{\infty}f(k)\leq f(1)+\int_{1}^{\infty}f(x)dx$.

We bound the second sum in (\ref{PASS.eq}),
\begin{align*}
\lefteqn{\sum_{n_{1}=-\infty}^{-1}\sum_{l=-n_{1}+1}^{\infty}P_{2}(n_{1},l)}\\
 & =\sum_{n_{1}=-\infty}^{-1}\sum_{l=-n_{1}+1}^{\infty}\frac{8\sqrt{-n_{1}+1}}{\pi(l+n_{1}-2)^{3/2}}\exp\left(-\frac{\upsilon(l)^{2}}{2(-n_{1}+1)}\right)
\end{align*}
We can ignore the small constants and write
\begin{align*}
P & =\sum_{n_{1}=-\infty}^{-1}\sum_{l=-n_{1}+1}^{\infty}\frac{8}{\pi(l+n_{1})^{3/2}}\exp\left(-\frac{\upsilon(l)^{2}}{2(-n_{1})}\right)\sqrt{-n_{1}}\\
 & =\sum_{n_{1}=-\infty}^{-1}\sum_{l=-n_{1}+1}^{\infty}\frac{8\sqrt{-n_{1}}}{\pi(l+n_{1})^{3/2}}2^{-\frac{l\tau}{-n_{1}}}l^{-\frac{3l}{-n_{1}}}\\
 & \leq\int_{1}^{\infty}\int_{t}^{\infty}\frac{8\sqrt{t}}{\pi(l-t)^{3/2}}2^{-\frac{l\tau}{t}}l^{-\frac{3l}{t}}dldt\\
 & =\int_{1}^{\infty}\int_{1}^{\infty}\frac{8\sqrt{t}}{\pi t^{3/2}(\tilde{l}-1)^{3/2}}2^{-\tilde{l}\tau}\tilde{l}^{-3\tilde{l}}t^{-3\tilde{l}}td\tilde{l}dt\\
 & =\int_{1}^{\infty}\int_{1}^{\infty}\frac{8}{\pi(\tilde{l}-1)^{3/2}}2^{-\tilde{l}\tau}\tilde{l}^{-3\tilde{l}}t^{-3\tilde{l}}d\tilde{l}dt\\
 & =\int_{1}^{\infty}\left(\int_{1}^{\infty}t^{-3\tilde{l}}dt\right)\frac{8}{\pi(\tilde{l}-1)^{3/2}}2^{-\tilde{l}\tau}\tilde{l}^{-3\tilde{l}}d\tilde{l}\\
 & =\int_{1}^{\infty}\frac{1}{3\tilde{l}-1}\frac{8}{\pi(\tilde{l}-1)^{3/2}}2^{-\tilde{l}\tau}\tilde{l}^{-3\tilde{l}}d\tilde{l}\\
 & =2^{-\tau}\int_{1}^{\infty}\frac{1}{3\tilde{l}-1}\frac{8}{\pi(\tilde{l}-1)^{3/2}}2^{-(\tilde{l}-1)\tau}\tilde{l}^{-3\tilde{l}}d\tilde{l}
\end{align*}
The remaining integral is clearly convergent, and decreasing in $\tau$.
Therefore $P\leq K2^{-\tau}$
\end{IEEEproof}
There are two important implications of Theorem \ref{Upper.thm}.
First is that for $\tau$ sufficiently large, $P_{A}(X_{n})<1$, and
in fact $P_{A}(X_{n})$ can be made arbitrarily small for large enough
$\tau$. This is an important theoretical validation of Definition
\ref{atypdef.thm} and the resulting criterion (\ref{Diid.eq}) and
(\ref{approxcrit.eq}). If the theory had resulted in $P_{A}(X_{n})=1$
then everything would be atypical, and atypicality would be meaningless.
That this is not trivially satisfied is shown by Proposition \ref{Lower.thm}
just below. What that Proposition says is that if in the above equation
instead of $\frac{3}{2}\log l$ we had had $\frac{1}{2}\log l$, then
everything would have been atypical. Now, $\frac{1}{2}\log l$ corresponds
to ``forgetting'' that the length of an atypical sequence also needs
to be encoded for the resulting sequence to be decodable. Thus, it
is the strict adherence to decodability that has lead to a meaningful
criterion. So, although decodability at first seems unrelated to detection,
it turns out to be of crucial importance. Similarly, at first the
term $\frac{3}{2}\log l$ may have seen arbitrary. However, this is
just (within a margin) sufficient to ensure that not everything becomes
atypical.

\begin{figure}[tbh]
\begin{centering}
\includegraphics[width=3.5in]{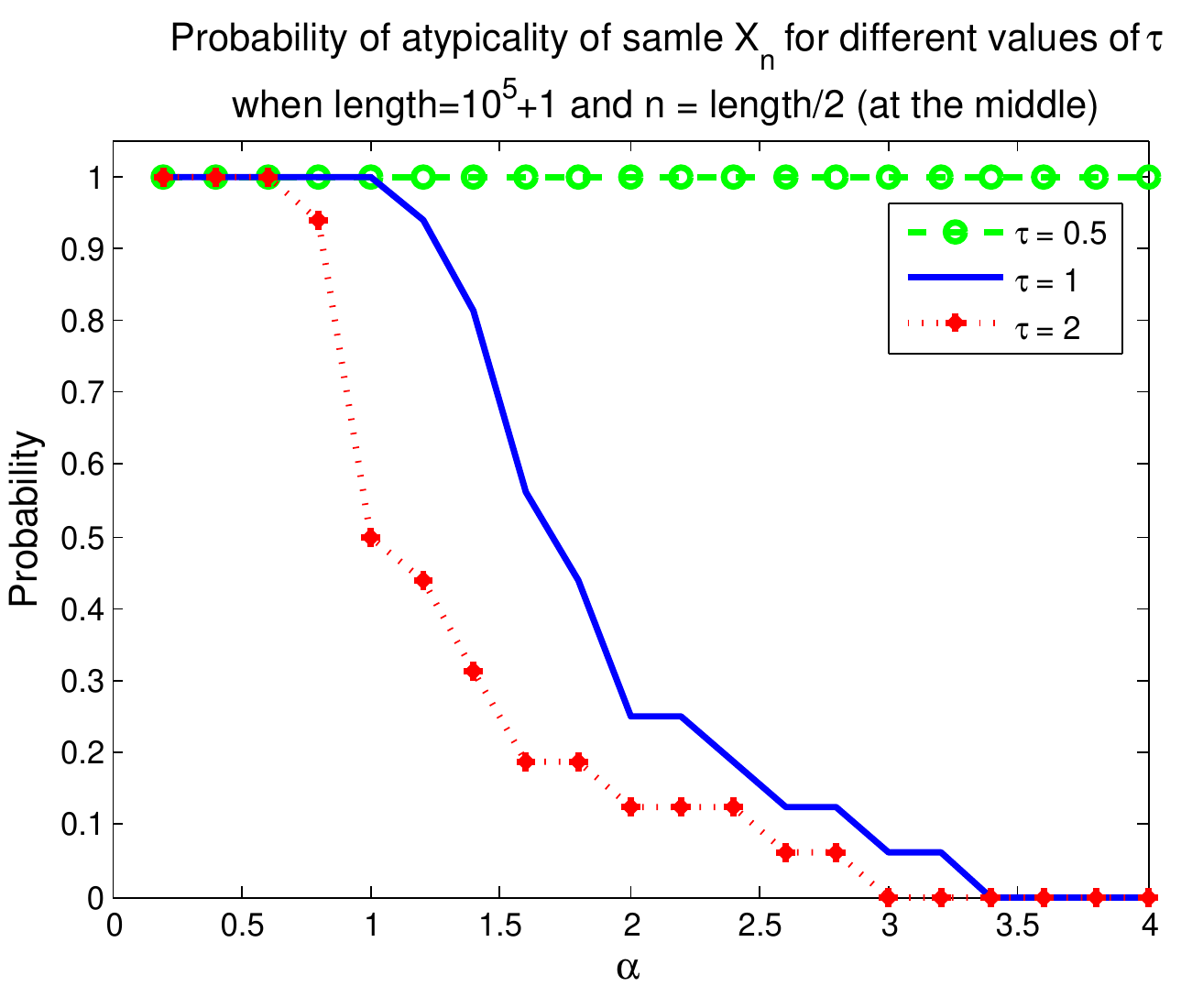}
\par\end{centering}
\caption{\label{Transition.fig}Transition between divergence and convergence
as a function of $\alpha$}

\end{figure}

The second important implication of Theorem \ref{Upper.thm} is that
it validates the meaning of $\tau$. The way we introduced $\tau$
was as the number of bits needed to encode the fact that an atypical
sequence starts, and therefore we should put $\tau=-\log P(\mbox{atypical sequence starts})$.
Theorem \ref{Upper.thm} confirms that $\tau$ has the desired meaning
for purely random sequences. And the reasons is this is not trivial
is that $\tau$ was chosen from the probability of an atypical \emph{sequence},
while Theorem \ref{Upper.thm} gives the probability of a \emph{sample}
being atypical.
\begin{prop}
\label{Lower.thm}Consider the case $p=\frac{1}{2}$. Suppose instead
of (\ref{CLT.eq}) we use the criterion 
\begin{align}
\left|\frac{\sum_{i=1}^{l}X_{i}-p}{\sqrt{\frac{1}{4}l}}\right| & >\sqrt{2\tau\ln2+\alpha\ln l}\label{CLT.eq-1}
\end{align}
(with $\alpha=3$ giving (\ref{CLT.eq})). Then if $\alpha\leq1$,
the probability that a given sample $X_{n}$ is part of an atypical
subsequence is $P_{A}(X_{n})=1$.
\end{prop}
\begin{IEEEproof}
We can assume that $n=0$. We will continue with the random walk framework
from the proof of Theorem \ref{Upper.thm}. Define the event
\begin{align*}
\bar{A}_{l} & =\left\{ \sum_{i=-l+1}^{0}(2X_{i}-1)>\sqrt{l\ln2(2\tau+\alpha\log l)}\right\} \\
\underbar{A}_{l} & =\left\{ \sum_{i=-l+1}^{0}(2X_{i}-1)<\sqrt{l\ln2(2\tau+\alpha\log l)}\right\} 
\end{align*}
and
\begin{align*}
\upsilon(l)= & \left\lceil \sqrt{l\ln2(2\tau+\alpha\log l)}\right\rceil 
\end{align*}
Then 
\[
P_{A}(X_{0})\geq P\left(\bigcup_{l=0}^{\infty}\bar{A}_{l}\cup\bigcup_{l=0}^{\infty}\underbar{A}_{l}\right)
\]
Namely, we declare that $X_{0}$ is atypical if it is the endpoint
of an atypical sequence $\{x[-l],x[-l+1],\ldots,x[0]\}$ for some
$l$. Clearly, $X_{0}$ could be the start or midpoint of an atypical
sequence, so this a rather loose lower bound. Now we can write
\begin{align*}
\lefteqn{P\left(\bigcup_{l=0}^{\infty}\bar{A}_{l}\cup\bigcup_{l=0}^{\infty}\underbar{A}_{l}\right)}\\
 & =1-P\left(\bigcap_{l=0}^{\infty}\bar{A}_{l}^{c}\cap\bigcap_{l=0}^{\infty}\underbar{A}_{l}^{c}\right)\\
 & =1-\prod_{l=0}^{\infty}P\left(\bar{A}_{l}^{c}\cap\underbar{A}_{l}^{c}\left|\bigcap_{k=0}^{l-1}\bar{A}_{k}^{c}\cap\bigcap_{k=0}^{l-1}\underbar{A}_{k}^{c}\right.\right)\\
 & =1-\prod_{l=0}^{\infty}\left[1-P\left(\bar{A}_{l}\cup\underbar{A}_{l}\left|\bigcap_{k=0}^{l-1}\bar{A}_{k}^{c}\cap\bigcap_{k=0}^{l-1}\underbar{A}_{k}^{c}\right.\right)\right]
\end{align*}
Consider the probability $P\left(\bar{A}_{l}\cup\underbar{A}_{l}\left|\bigcap_{k=0}^{l-1}\bar{A}_{k}^{c}\cap\bigcap_{k=0}^{l-1}\underbar{A}_{k}^{c}\right.\right)$.
The only way the conditional event can happen is if $S_{l-1}=\upsilon(l)-1$
and $X_{l}=1$ or $S_{l-1}=-\upsilon(l)+1$ and $X_{l}=-1$. Here
we have
\begin{align*}
\lefteqn{P\left(S_{l-1}=\upsilon(l)-1\right)}\\
 & =N_{l-1}(0,\upsilon(l)-1)2^{-l+1}\\
 & =\left(\begin{array}{c}
l-1\\
\frac{1}{2}(l+\upsilon(l)-2)
\end{array}\right)2^{-l+1}\\
 & \geq\sqrt{\frac{l-1}{2(l+\upsilon(l)-2)(l-\upsilon(l))}}2^{(l-1)H\left(\frac{\frac{1}{2}(l+\upsilon(l)-2)}{l-1}\right)}2^{-l+1}\\
 & =\sqrt{\frac{l-1}{2((l-1)^{2}-(\upsilon(l)-1)^{2})}}2^{(l-1)H\left(\frac{\frac{1}{2}(l+\upsilon(l)-2)}{l-1}\right)}2^{-l+1}\\
 & \geq\sqrt{\frac{1}{2l}}2^{(l-1)H\left(\frac{\frac{1}{2}(l+\upsilon(l)-2)}{l-1}\right)}2^{-l+1}
\end{align*}
Here
\begin{align*}
\lefteqn{(l-1)H\left(\frac{\frac{1}{2}(l+\upsilon(l)-2)}{l-1}\right)-l+1}\\
= & (l-1)\left(H\left(\frac{1}{2}+\frac{1}{2}\frac{\upsilon(l)-1}{l-1}\right)-1\right)\\
\geq & -\frac{2}{\ln2}(l-1)\left(\frac{1}{2}\frac{\upsilon(l)-1}{l-1}\right)^{2}+(l-1)o\left(\left(\frac{\upsilon(l)-1}{l-1}\right)^{3}\right)\\
= & -\frac{1}{2\ln2}\frac{\left(\upsilon(l)-1\right)^{2}}{l-1}+\frac{(\upsilon(l)-1)^{3}}{(l-1)^{2}}\epsilon\left(\frac{1}{l}\right)\\
\geq & -\frac{1}{2\ln2}\left(\frac{\upsilon(l)^{2}}{l}+\frac{\upsilon(l)^{2}}{(l-1)^{2}}+\frac{1}{l-1}\right)+\frac{\upsilon(l){}^{3}}{l^{2}}\epsilon\left(\frac{1}{l}\right)\\
= & -\frac{1}{2\ln2}\frac{\upsilon(l)^{2}}{l}+\frac{\upsilon(l){}^{3}}{l^{2}}\epsilon\left(\frac{1}{l}\right)\\
\geq & -\frac{1}{2\ln2}\frac{\upsilon(l)^{2}}{l}-\frac{\upsilon(l){}^{3}}{l^{2}}
\end{align*}
Where the last inequality is only true for $l$ sufficiently large,
as for some $l_{0}$ we have $\forall l>l_{0}:|\epsilon(l^{-1})|<1$.
Then
\begin{align*}
P\left(S_{l-1}=\upsilon(l)-1\right) & \geq\sqrt{\frac{1}{2l}}2^{-\tau}l^{-\frac{\alpha}{2}}2^{-\frac{\upsilon(l){}^{3}}{l^{2}}}\\
 & =\sqrt{\frac{1}{2}}2^{-\tau}l^{-\frac{\alpha+1}{2}}2^{-\frac{\upsilon(l){}^{3}}{l^{2}}}
\end{align*}
And
\begin{align*}
\lefteqn{\ln\left(1-P\left(\bigcup_{l=0}^{\infty}\bar{A}_{l}\cup\bigcup_{l=0}^{\infty}\underbar{A}_{l}\right)\right)}\\
\leq & \sum_{l=1}^{\infty}\ln\left(1-\sqrt{\frac{1}{2}}2^{-\tau}l^{-\frac{\alpha+1}{2}}2^{-\frac{\upsilon(l){}^{3}}{l^{2}}}\right)\\
\leq & \sum_{l=0}^{\infty}-\sqrt{\frac{1}{2}}2^{-\tau}l^{-\frac{\alpha+1}{2}}2^{-\frac{\upsilon(l){}^{3}}{l^{2}}}
\end{align*}
Here $\lim_{l\to\infty}\frac{\upsilon(l){}^{3}}{l^{2}}=0$. So, for
example, for $l$ sufficiently large, $\frac{\upsilon(l){}^{3}}{l^{2}}\leq1$.
Then
\begin{align*}
\lefteqn{\ln\left(1-P\left(\bigcup_{l=0}^{\infty}\bar{A}_{l}\cup\bigcup_{l=0}^{\infty}\underbar{A}_{l}\right)\right)}\\
\leq & \sum_{l=l_{0}}^{\infty}-\frac{1}{2}\sqrt{\frac{1}{2}}2^{-\tau}l^{-\frac{\alpha+1}{2}}
\end{align*}
This is divergent for $\alpha\leq1$ proving that $P\left(\bigcup_{l=0}^{\infty}\bar{A}_{l}\cup\bigcup_{l=0}^{\infty}\underbar{A}_{l}\right)=1$.
\end{IEEEproof}
Theorem \ref{Upper.thm} states that for $\alpha=\frac{3}{2}$ $P_{A}(X_{n})<1$
(convergence), while Proposition \ref{Lower.thm} shows that for $\alpha=\frac{1}{2}$
$P_{A}(X_{n})=1$ (divergence). There is a gap between those values
of $\alpha$ that is hard to fill in theoretically. We have therefore
tested it out numerically, see Fig. \ref{Transition.fig}. Of course,
testing convergence numerically is not quite well-posed. Still the
figure indicates that the phase transitions between divergence and
convergence happens right around $\alpha=1$.

\subsection{Recursive coding}

\label{recursive.sec}Instead of using Definition \ref{atypdef.thm}
directly, we could approach the problem as follows. First the sequence
is encoded with the typical code. Now, if the distribution of the
sequence is in agreement with the typical code, the results should
be a sequence of iid binary bits with $P(X_{i}=1)=\frac{1}{2}$ \cite{CoverBook},
i.e., a purely random sequence; and this sequence cannot be further
encoded. We can now try if we can further encode the sequence with
a (universal) code. If so, we categorize the sequence as atypical.
Let $l^{*}$ be the length of the sequence after typical coding. In
(\ref{LtLaiid.eq}) the typical and atypical codelengths are therefore
\begin{align}
L_{t} & =l^{*}-\log\epsilon\nonumber \\
L_{a} & =l^{*}H(\hat{p}^{*})+\frac{3}{2}\log l^{*}-\log\epsilon+\tau^{*}\label{LtLaiidr.eq}
\end{align}
Here $\hat{p}^{*}$ is the estimated $p$ for the \emph{encoded} sequence.
Now
\begin{align}
l^{*} & =l\left(\hat{p}\log\frac{1}{p}+(1-\hat{p})\log\frac{1}{1-p}\right)\\
l^{*}H(\hat{p}^{*}) & \sim lH(\hat{p})\label{LHapprox.eq}\\
\log l^{*} & =\log l+\log\left(\hat{p}\log\frac{1}{p}+(1-\hat{p})\log\frac{1}{1-p}\right)\\
 & \sim\log l
\end{align}
The argument for (\ref{LHapprox.eq}) is as follows (without doing
detailed calculations): if we encode a sequence with a ``wrong''
code and then later re-encode with the ``correct'' code (for the
induced statistic), the result is the same as originally encoding
with the correct code. Thus the criterion (\ref{LtLaiid.eq}) and
(\ref{LtLaiidr.eq}) are \emph{approximately} equivalent. We can state
this as follows
\begin{prop}
Definition \ref{atypdef.thm} can be applied to encoded sequences
instead of the original data.
\end{prop}
This of course ignores all integer constraints, block boundaries etc.
But the importance of this statement is that it is sometimes easier
to operate on (partially) encoded sequences simply because the amount
of data has already been reduced, and the problem has been standardized:
as such, we do not need to know the typical codebook or even the model
of typical data since everything under the typical model has been
reduced to a stream of iid binary digits, and atypicality algorithms
can therefore be applied to data streams without knowledge of what
is the original data. It also means that theoretical results such
as Theorem \ref{Upper.thm} where we assume typical data is iid uniform
has general applicability.

However, first encoding the sequence and then doing atypicality detection
also has disadvantages in a practical, finite length setting. Atypical
subsequences become embedded in typical sequences in unpredictable
ways. For example, it could be difficult to determine where exactly
an atypical sequence starts and ends. Our practical implementation
therefore uses Definition \ref{atypdef.thm} directly.

\section{General Case}

Return to the problem considered at the start of Section \ref{binaryiid.sec}
where we are given a sequence $x$ of fixed length $l$ and we need
to determine if it is atypical. In the iid case this is a simple hypothesis
test problem and the solution is given by (\ref{atyphyp.eq}). In
the general case we would like find to alternative explanations from
a large abstract class of models. The issue is that it is often possible
to fit an alternative model very well to the data if we just allow
complex enough models \textendash{} the well known Occam's razor problem
\cite{CoverBook}. Rissanen's MDL \cite{Rissanen78,Rissanen83,Rissanen84}
is a solution to this problem. Therefore, in the general case, even
for fixed length sequences, the problem is not a straightforward hypothesis
test problem, and we have to resort to information theory.

\subsection{\label{FSM.sec}Finite State Machines}

On possible class of models in the general case is the class of finite
state machines (FSM). Rissanen \cite{Rissanen86b} defines the complexity
of a sequence $x^{l}$ in the class of FSM sources by
\begin{equation}
I(x^{l})=\min\{-\log\hat{P}(x^{l}|f_{j})+\log^{*}j+c\}\label{FSMcomplexity.eq}
\end{equation}
where $f_{1},f_{2},\ldots$ is a sequence of state machines, and where
we have used $\hat{P}(x^{l}|f_{j})$ to emphasize that the probability
is estimated. Rissanen uses Laplace's estimator, but the KT-estimator
\cite{WillemsAl95,CoverBook} could also be used. Except for integer
constraints, this is a valid descriptive length, and can therefore
be used in Definition \ref{atypdef.thm}. This is a natural extension
of the iid case considered in Section \ref{binaryiid.sec}. As opposed
to Kolmogorov complexity, this complexity could actually be calculated,
although with high complexity. Because of the complexity, it is mostly
useful for theoretical considerations, and one result is the following
generalization of Theorem \ref{1storderupper.thm}
\begin{thm}
Assume that the typical distribution is iid uniform. If the atypical
descriptive length is given by (\ref{FSMcomplexity.eq}) with a maximum
number of states independent of $l$, the probability of an intrinsically
atypical sequence $P_{A}(l)$ satisfies 
\begin{equation}
\lim_{l\to\infty}\frac{\ln P_{A}(l)}{-\frac{3}{2}\ln l}=1\label{FSMlower.eq}
\end{equation}
\end{thm}
\begin{IEEEproof}
Since we consider all state machines with the number of states up
to a certain maximum, this must also include the state machine with
a single state. This is equivalent to the iid model in Section \ref{binaryiid.sec},
and we therefore get the lower bound in (\ref{FSMlower.eq}). The
proof will be to upper bound the probability. As in Section \ref{binaryiid.sec}
we use $\log l+\tau$ bits to indicate beginning and end of atypical
sequences. The probability that a sequence $x^{l}$ is atypical therefore
is 
\begin{align*}
P_{A}(l) & =P(I(x^{l})+\log l+\tau>l)\\
 & =P(\bigcup_{f_{j}}-\log\hat{P}(x^{l}|f_{j})+\log^{*}j+c+\log l+\tau>l)\\
 & \leq\bigcup_{f_{j}}P(-\log\hat{P}(x^{l}|f_{j})+\log l+\tau>l)
\end{align*}
We will prove that $P(-\log\hat{P}(x^{l}|f_{j})+\log l+\tau>l)\leq K_{j}l^{-(k+2)/2}$
for constants $K_{j}$ and $k$ the number of states in the state
machine, and since the slowest decay dominates, we get the upper bound
for (\ref{FSMlower.eq}).

For a fixed state machine $f$ the code length according to \cite[(3.6)]{Rissanen86b}
is
\[
L(x^{l}|f)=\sum_{s}\log\left(\begin{array}{c}
n_{s}(x^{l})\\
n_{0|s}(x^{l})
\end{array}\right)+\sum_{s}\log(n_{s}(x^{l})+1)
\]
where $n_{s}(x^{l})$ denotes the number of occurrences of state $s$
in $x^{l}$ and $n_{0|s}(x^{l})$ the number of times the next symbols
is $0$ at this state. Further, from \cite[13.2]{CoverBook}
\begin{align}
L(x^{l}|f) & \geq\sum_{s}n_{s}(x^{l})H\left(\frac{n_{0|s}(x^{l})}{n_{s}(x^{l})}\right)-\frac{1}{2}\log n_{s}(x^{l})\nonumber \\
 & -\frac{1}{2}\log\left(8\frac{n_{0|s}(x^{l})}{n_{s}(x^{l})}\frac{n_{1|s}(x^{l})}{n_{s}(x^{l})}\right)\nonumber \\
 & +\log(n_{s}(x^{l})+1)\label{Lf.eq}
\end{align}
We want to upper bound the probability of the event $L(x^{l}|f)+\log l+\tau<l$.
We can write
\[
\log(n_{s}(x^{l})+1)-\frac{1}{2}\log(n_{s}(x^{l}))=\frac{1}{2}\log l+\log\left(\frac{1}{l}\frac{\left(n_{s}(x^{l})+1\right)^{2}}{n_{s}(x^{l})}\right).
\]
Let 
\[
r(x^{l})=\sum_{s}n_{s}(x^{l})\left(H\left(\frac{n_{0|s}(x^{l})}{n_{s}(x^{l})}\right)-1\right)
\]
and let $R(x^{l})$ be the remaining small terms in (\ref{Lf.eq})
dependent on $x^{l}$,
\begin{align*}
R(x^{l}) & =\sum_{s}-\frac{1}{2}\log\left(8\frac{n_{0|s}(x^{l})}{n_{s}(x^{l})}\frac{n_{1|s}(x^{l})}{n_{s}(x^{l})}\right)\\
 & +\log\left(\frac{1}{l}\frac{\left(n_{s}(x^{l})+1\right)^{2}}{n_{s}(x^{l})}\right).
\end{align*}
Then we have to upper bound (notice that $\sum_{s}n_{s}(x^{l})=l$),
\[
P\left(-r(x^{l})-R(x^{l})\geq\tau+\frac{k+2}{2}\log l\right)
\]
The Chernoff bound is
\begin{align*}
\lefteqn{P\left(-r(x^{l})-R(x^{l})\geq\tau+\frac{k+2}{2}\log l\right)}\\
 & \leq\exp(-t(\tau+\frac{k+2}{2}\log l))M(t)
\end{align*}
or
\begin{align*}
\lefteqn{\ln P\left(-r(x^{l})-R(x^{l})\geq\tau+\frac{k+2}{2}\log l\right)}\\
 & \leq-t(\tau+\frac{k+2}{2}\log l)+\ln M(t)
\end{align*}
where
\[
M(t)=E\left[\exp(-t(r(x^{l})+R(x^{l}))\right]
\]
In order to get a valid bound, we need to show that $M(t)<K<\infty$
independent of $l$ for $t<\ln2$. Now it's easy to see that $\exp(-tR(x^{l}))\leq K<\infty$
for all $t$ and $l$. So, we have to show
\[
E\left[\exp(-t(r(x^{l}))\right]\leq K<\infty
\]
We have to show that this is true for all state machines in the class
of finite state machines with $k$ states, which can be done by showing
\[
\max_{\mbox{FSM with }k\mbox{ states}}E\left[\exp(-t(r(x^{l}))\right]\leq K
\]
It turns out it easier to prove this if we expand the class over which
we take the maximum, and clearly expanding the class does not decrease
the maximum. A FSM with $k$ states is a function $f(x^{l})\in\{1,\ldots,k\}$
that satisfies that if $f(x^{m})=f(\tilde{x}^{m})=s$ then $f(x^{m}b)=f(\tilde{x}^{m}b)$
for any bit $b$ \cite{Rissanen86b}, i.e., if the FSM is in state
$s$ after $m$ steps, the next state transition is only dependent
on the next bit, not how it got to state $s$. We extend the class
by dispensing with this requirement. We can then describe the 'program'
we run as follows. Based on $x^{m}$ we choose a state $s_{m}\in\{1,\ldots,k\}$
\emph{without} having any knowledge about $x_{m+1}$, except that
it is independent and uniformly distributed (by the assumption on
the typical distribution). We can think of this slightly differently.
The program puts $x_{m+1}$ into bucket $s\in\{1,\ldots,k\}$ and
updates $n_{s}(m)$ and $n_{i|s}(m)$, in order to maximize $E\left[\exp(-t(r(x^{l}))\right]$.
It does so based on past data $x^{m}$. Now, as opposed to the state
machine setup, the choice of $s_{m}$ in no way restricts the choices
of states (or buckets) $s_{n}$, $n>m$. Since the program has no
knowledge of $x_{m+1}$ the program cannot optimize $s_{m}$ based
on the \emph{values} of $x^{m}$. Rather, it is sufficient to look
at $n_{s}(m)$. It is now easy to see that the worst case is obtained
if the bits are distributed evenly in the states. Thus, the worst
case of $r(x^{l})$ is 
\[
r(x^{l})=\sum_{s}\frac{l}{k}\left(H\left(\frac{n_{0|s}(x^{l})}{l/k}\right)-1\right)
\]
where the $n_{0|s}(x^{l})$ are independent of $s$. Thus, the problem
is reduced to the case of a single state, which is showing that
\begin{equation}
E\left[\exp\left(tl\left(1-H\left(\frac{k}{l}\right)\right)\right)\right]\leq K<\infty\label{rlfin1.eq}
\end{equation}
Here we have
\begin{align*}
\lefteqn{E\left[\exp\left(tl\left(1-H\left(\frac{k}{l}\right)\right)\right)\right]}\\
 & =\sum_{k=0}^{l}2^{\frac{t}{\ln2}l(1-H(\frac{k}{l}))}\left(\begin{array}{c}
k\\
l
\end{array}\right)2^{-l}\\
 & =1+2\sum_{k=1}^{l/2}2^{\frac{t}{\ln2}l(1-H(\frac{k}{l}))}\left(\begin{array}{c}
k\\
l
\end{array}\right)2^{-l}\\
 & \leq1+2\sum_{k=1}^{l/2}2^{\frac{t}{\ln2}l(1-H(\frac{k}{l}))}2^{-l}2^{lH(\frac{k}{l})}\sqrt{\frac{l}{\pi k(l-k)}}\\
 & =1+2\sum_{k=1}^{l/2}2^{\left(1-\frac{t}{\ln2}\right)l(1-H(\frac{k}{l}))}\sqrt{\frac{l}{\pi k(l-k)}}\\
 & \leq1+2\sum_{k=1}^{l/2}2^{-\left(1-\frac{t}{\ln2}\right)\frac{2}{\ln2}l\left(\frac{k}{l}-\frac{1}{2}\right)^{2}}\sqrt{\frac{l}{\pi k(l-k)}}
\end{align*}
where we have used \cite[13.2]{CoverBook} and (\ref{Hbound.eq}).
The sum is actually decreasing as a function of $l$, but this seems
hard to prove. Instead we upper bound the sum by

\begin{align*}
\lefteqn{\sum_{k=1}^{l/2}2^{-\left(1-\frac{t}{\ln2}\right)\frac{2}{\ln2}l\left(\frac{k}{l}-\frac{1}{2}\right)^{2}}\sqrt{\frac{l}{\pi k(l-k)}}}\\
\leq & \int_{1}^{l/2}2^{-\left(1-\frac{t}{\ln2}\right)\frac{2}{\ln2}l\left(\frac{k}{l}-\frac{1}{2}\right)^{2}}\sqrt{\frac{l}{\pi k(l-k)}}dk
\end{align*}
Here we can upper bound $\sqrt{\frac{l}{\pi k(l-k)}}\leq\frac{\frac{4k\left(\sqrt{\frac{1}{l}}-1\right)}{k}+2}{\sqrt{\pi}}$
for $1\leq k\leq\frac{l}{2}$. Then
\begin{align*}
\lefteqn{\int_{1}^{l/2}2^{-\left(1-\frac{t}{\ln2}\right)\frac{2}{\ln2}l\left(\frac{k}{l}-\frac{1}{2}\right)^{2}}\sqrt{\frac{l}{\pi k(l-k)}}dk}\\
 & \leq\int_{1/l}^{1/2}2^{-\left(1-\frac{t}{\ln2}\right)\frac{2}{\ln2}l\left(x-\frac{1}{2}\right)^{2}}\frac{4x\left(\sqrt{\frac{1}{l}}-1\right)+2}{\sqrt{\pi}}ldx\\
 & \leq\int_{-\infty}^{1/2}2^{-\left(1-\frac{t}{\ln2}\right)\frac{2}{\ln2}l\left(x-\frac{1}{2}\right)^{2}}\frac{4x\left(\sqrt{\frac{1}{l}}-1\right)+2}{\sqrt{\pi}}ldx\\
 & =\frac{K_{1}}{\sqrt{l}}+K_{2}
\end{align*}
for some constants $K_{1},K_{2}$, using Gaussian moments. This proves
(\ref{rlfin1.eq}).

\end{IEEEproof}
\begin{figure}[tbh]
\begin{centering}
\includegraphics[width=3.5in]{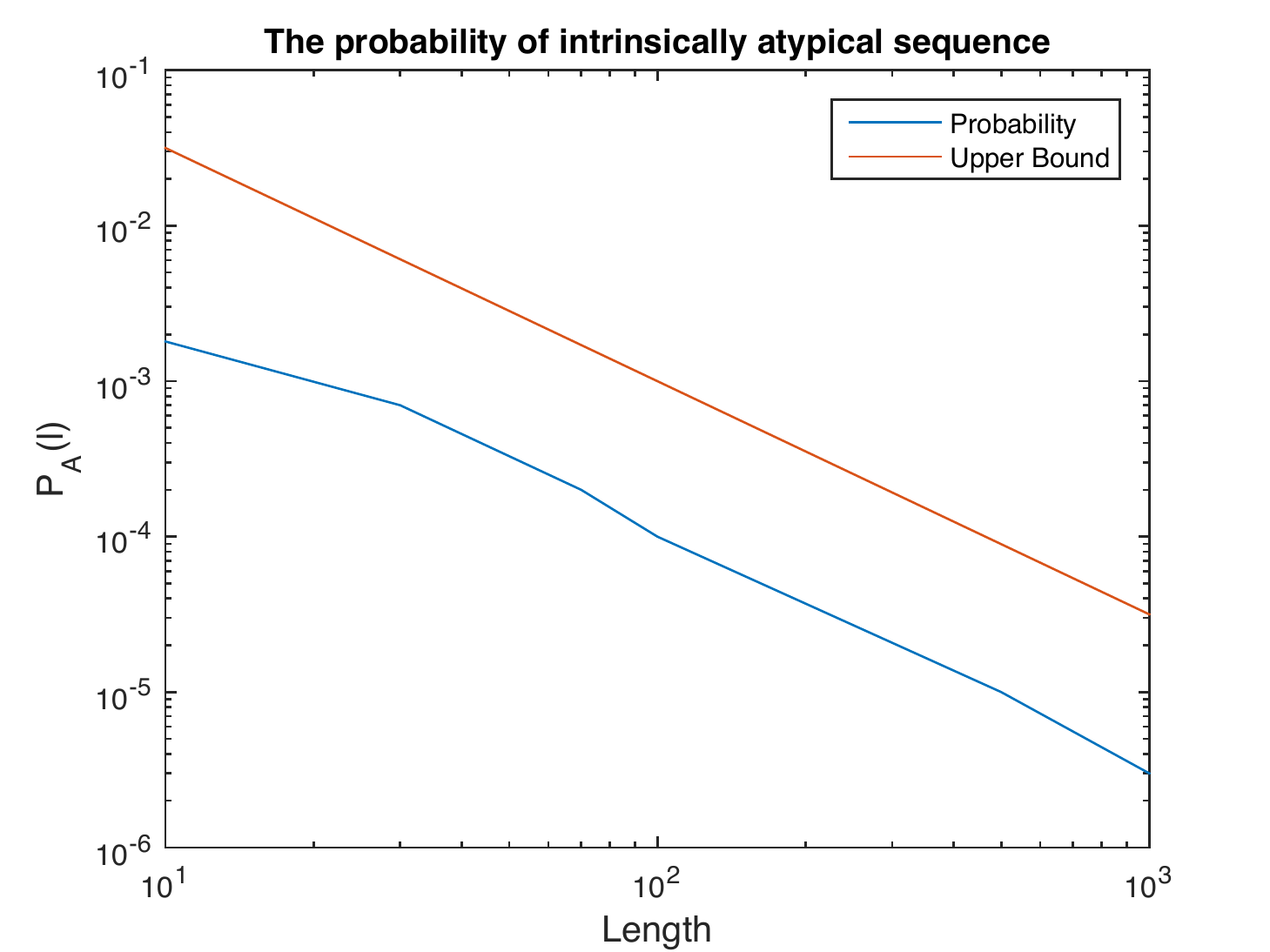}
\par\end{centering}
\caption{\label{IntrinsicCTW.fig}Probability of an intrinsically atypical
sequence. The typical distribution is iid uniform, and for detection
of atypical sequences the CTW algorithm has been used (Section \ref{Universal.sec}).}

\end{figure}

While the Theorem is for the typical model iid uniform, as outlined
in Section \ref{recursive.sec} in principle it also applies to general
sources, since we can first encode and then look for atypical sequences.

The theorem shows that looking for more complex explanations for data
does not essentially increase the probability of intrinsically atypical
sequences. Fig. \ref{IntrinsicCTW.fig} (compare with Fig. \ref{PA.fig})
confirms this experimentally. The atypical detection is based on CTW,
which as explained in Section \ref{Universal.sec} below, is a good
approximation of FSM modeling. On the other hand, if one of the FSM
models do in fact fit the data, the chance of detecting the sequence
is greatly increased, although hard to quantify. If we think of intrinsically
atypical sequences as false alarms, this shows the power of the methodology.

Since FSM sources has the same $P_{A}(l)$ as in the iid case, it
seems reasonable to conjecture that Theorem \ref{Upper.thm} is still
valid, that is $P_{A}(X_{n})<1$ for sufficiently large $\tau$, which
is clearly an essential theoretical property of atypicality. However,
as Theorem \ref{Upper.thm} does not follow directly from Theorem
\ref{1storderupper.thm}, to verify the conjecture requires a formal
proof which we do not have at present.

\subsection{\label{Universal.sec}Atypical Encoding}

In terms of coding, Definition \ref{atypdef.thm} can be stated in
the following form
\[
C(x|\mathcal{P})-C(x)>0
\]
Here $C(x|\mathcal{P})$ is the code length of $x$ encoded with the
optimum coder according to the typical law, and $C(x)$ is $x$ encoded
'in itself.' As argued in Section \ref{binaryiid.sec}, we need to
put a 'header' in atypical sequences to inform the encoder that an
atypical encoder is used. We can therefore write $C(x)=\tau+\tilde{C}(x)$,
where $\tau$ is the number of bits for the 'header,' and $\tilde{C}(x)$
is the number of bits used for encoding the data itself. For encoding
the data itself an obvious solution is to use a universal source coder.
There are many approaches to universal source coding: Lempel-Ziv \cite{CoverBook,ZivLempel77,ZivLempel78},
Burrows-Wheeler transform \cite{EffrosVerduAl02}, partial predictive
mapping (PPM) \cite{ClearyWitten84,Moffat90}, or T-complexity \cite{Titchener98,HamanoYamamoto08,KawaharadaSpeidelAl05,Speidel09,SpeidelAl07,SpeidelGulliver12,YangSpeidel05},
and anyone of them could be applied to the problem considered in this
paper. The idea of atypicality is not linked to any particular coding
strategy. In fact a coding strategy does not need to be decided. We
could try several source coders and choose the the one giving the
shortest code length; or they could even be combined as in \cite{VolfWillems98}.
However, to control complexity, we choose a single source coder. The
most popular and simplest approach to source coding is perhaps Lempel-Ziv
\cite{CoverBook,ZivLempel77,ZivLempel78}. The issue with this is
that while it is optimum in the sense that $\limsup_{l\to\infty}\frac{C(x^{l})}{l}=H(X)\mbox{ wp }1$,
the convergence is very slow. According to \cite{JacquetSzpankowski11}
$E\left[\frac{C(x^{l})}{l}\right]-H(X)\sim\frac{1}{\log l}$ while
$\var\left[\frac{C(x^{l})}{l}\right]\sim\frac{1}{l}$. Thus, Lempel-Ziv
is poor for short sequences, which is exactly what we are interested
in for atypicality. 

We have therefore chosen to use the Context Tree Weighing (CTW) algorithm
\cite{WillemsAl95}. The CTW approach has some advantages in our setup:
it is a natural extension of the simple example considered in Section
\ref{binaryiid.sec}, it allows estimation of code length without
actually encoding, there is flexibility in how to estimate probabilities.
Importantly, it can be seen as a practical implementation of the FSM
based descriptive length used in Section \ref{FSM.sec}.

\subsection{\label{Typical.sec}Typical Encoding and Training}

In Definition \ref{atypdef.thm} and the example in Section \ref{binaryiid.sec}
we have assumed that the typical model of data is exactly known. If
that is the case, typical encoding is straightforward, using for example
arithmetic coding \textendash{} notice that we just need codelength,
which can be calculated for arithmetic coding without actually encoding.
However, in many cases the typical model is not known exactly. In
the simplest case, the typical model is from a small class parametrized
with a few parameters (e.g., binary iid with $P(X=1)$ unknown). In
that case, finding the typical model is a simple parameter estimation
problem, and we will not discuss this further. We will focus on the
case where the typical model is not given by a specific model. In
that case, it seems obvious to also use universal source coding for
typical data. However, this is not straightforward if we want to stay
faithful to the idea of Definition \ref{atypdef.thm} as we will argue
below.

Let $\mathcal{P}$ be the typical model and $\hat{\mathcal{P}}$
be the estimated typical model. When no specific model is given, in
principle $\hat{\mathcal{P}}$ could be given by an estimate of the
various joint probability mass functions. However, a more useful approach
is to estimate the conditional probabilities $p(x_{n}=1|x_{n-1},x_{n-2},\ldots,x_{n-N})$,
where $s=x_{n-1},x_{n-2},\ldots,x_{n-N}$ is called the context. If
the source has finite memory these probabilities characterize the
source, and otherwise they could give a good approximate model. The
issue is that there are $2^{N}$ possible contexts, so for $N$ even
moderately large the amount of training data required to even observe
every context is large, and to get good estimates for every context
it is even larger. Realistically, therefore not every probability
$p(x_{n}=1|s)$ can be estimated. This is an issue universal source
coding is designed to deal with, and we therefore turn to universal
source coding.

Let us assume we are given a single long sequence $t$ for training
\textendash{} rather than a model $\mathcal{P}$ \textendash{} and
based on this we need to encode a sequence $x$. Let us denote this
coder as $C(x|t)$. To understand what this means, we have to realize
that when $x$ is encoded according to $C(x|\mathcal{P})$ with a
known $\mathcal{P}$, the coding probabilities are \emph{fixed}; they
are not affected by $x$. As discussed in Section \ref{MDL.sec},
this is an important part of Definition \ref{atypdef.thm} that reacts
to 'outliers,' data that does not fit the typical model. If the coding
probabilities for $C(x|\mathcal{P})$ were allowed to depend on $x$,
in extreme case we would always have $C(x)=C(x|\mathcal{P})$.

The issue with universal source coders is that they often easily adapts
to new types of data, a desirable property of a good universal source
coder, but problematic in light of the above discussion. We therefore
need to 'freeze' the source coder, for example by not updating the
dictionary. However, because the training data is likely incomplete
as discussed above, the freezing should not be too hard. The resulting
encoder $C(x|t)$ is not a universal source coder, but rather a training
based fixed source coder, and the implementation can be quite different
from a universal source coder, requiring careful consideration.

We will suggest one algorithm based on the principle of the CTW algorithm.
This naturally complements using the CTW algorithm for atypical encoding,
but could also be used with other atypical encoders. The following
discussion requires a good knowledge of the CTW algorithm, which is
most easily obtained from \cite{WillemsAl97}.

The algorithm is based on estimating $P(x_{n}=1|s)$ for contexts
$s$. The estimate for a given $s$ is done with the KT-estimator
\cite{WillemsAl95,CoverBook} $P(x_{n}=1|s)=\frac{b_{s}+1/2}{a_{s}+b_{s}+1}$,
where $a_{s}$ and $b_{s}$ are the number of 0s and 1s respectively
seen with context $s$ in the training data $t$, but unaffected by
the test sequence $x$. As discussed above, the complication is that
not every context $s$ might be seen and that in particular long contexts
$s$ are rarely seen so that the estimates $P(X_{n}=1|s)$ might be
more accurate for shorter contexts. We solve that with the weighting
idea of the CTW algorithm; the weights can be thought of as a prior
distribution on different models. We can summarize this as follows.
For every context $s$, the subsequence associated with $s$ could
either be memoryless or it could have memory \cite{WillemsAl97}.
We call the former model $M_{1}$ and the latter $M_{2}$. The CTW
algorithm uses a prior distribution, weights, on the models $P(M_{1})=P(M_{2})=\frac{1}{2}$.
Our basic idea is to weigh with $P(M_{1}|t)$ and $P(M_{2}|t)$ instead
of $\frac{1}{2}$.

\begin{figure}[tbh]
\begin{centering}
\includegraphics[scale=0.75]{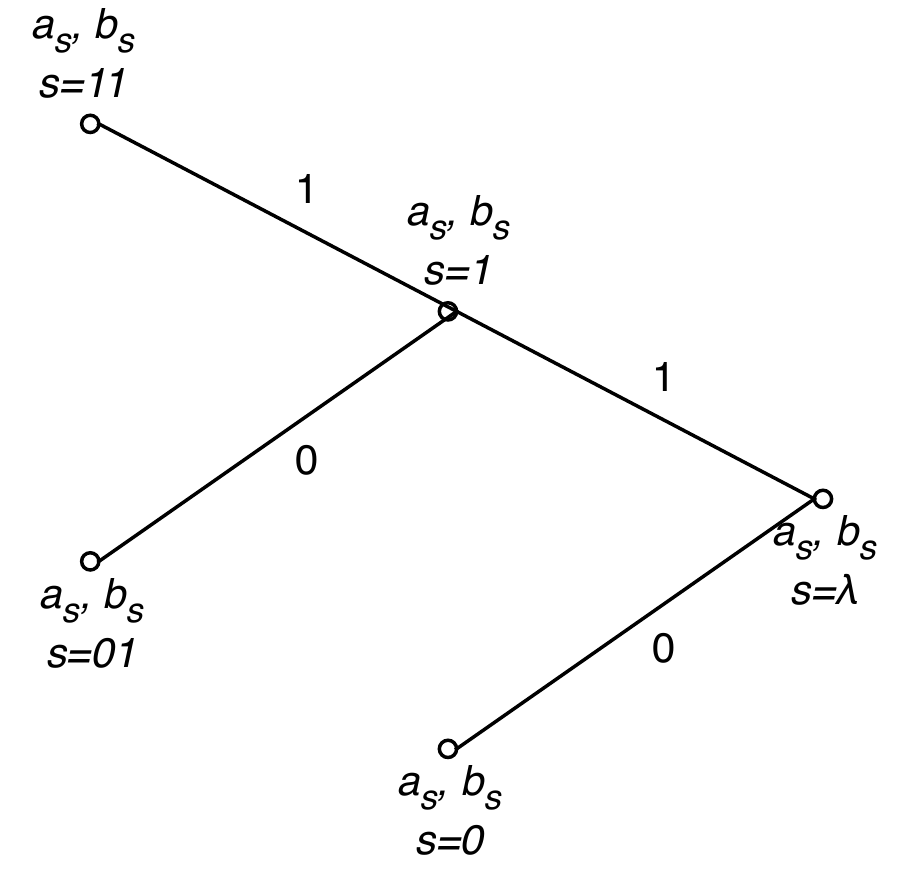}
\par\end{centering}
\caption{\label{ContextTree.fig}Example context tree.}

\end{figure}

The algorithm is best described through an example. We have trained
the algorithm with $t$, resulting in the context tree seen in Fig.
\ref{ContextTree.fig}. Now suppose we want to find a coding distribution
$P(1|010)$ for the actual data. We begin at the root and calculate
\begin{align*}
P(1|010,t) & =P(1|010,M_{1})P(M_{1}|t)+P(1|010,M_{2})P(M_{2}|t)\\
P(M_{i}|t) & =\frac{P(t|M_{i})P(M_{i})}{P(t)}=\frac{P(t|M_{i})}{P(t|M_{1})+P(t|M_{2})}\\
P(t|M_{1}) & =P_{e}(a_{s},b_{s})\\
P(t|M_{2}) & =P_{w}^{0s}(t)P_{w}^{1s}(t)\\
s & =\lambda\quad\mbox{(empty context)}
\end{align*}
under model $M_{1}$ the data is memoryless, so
\begin{align*}
P(1|010,M_{1}) & =P(1|M_{1})=\frac{b_{s}+1/2}{a_{s}+b_{s}+1};\quad s=\lambda
\end{align*}
To find $P(1|010,M_{2})$ we look in the $0$-node of the context
tree. Here we calculate similarly
\begin{align*}
P(1|010,t) & =P(1|010,M_{1})P(M_{1}|t)+P(1|010,M_{2})P(M_{2}|t)\\
P(M_{i}|t) & =\frac{P(t|M_{i})P(M_{i})}{P(t)}=\frac{P(t|M_{i})}{P(t|M_{1})+P(t|M_{2})}\\
P(t|M_{1}) & =P_{e}(a_{s},b_{s})\\
P(t|M_{2}) & =P_{w}^{0s}(t)P_{w}^{1s}(t)\\
s & =0
\end{align*}
and again under model $M_{1}$ the data is iid, so
\[
P(1|10,M_{1})=P(1|0,M_{1})=\frac{b_{s}+1/2}{a_{s}+b_{s}+1};\quad s=0
\]
and so on. From the context tree it is seen that the context $10$
has not been seen before; then the context $010$ has not been seen
either. Then we have 
\[
P(1|010,M_{2})=\frac{1}{2}.
\]
No more look-up is needed, and the calculation has completed.

The algorithm can be implemented as follows. We run the standard CTW
algorithm on the training data. To freeze, in each node corresponding
to the context $s$ we can pre-compute $P(M_{1}|t)$ and $P(1|s,M_{1})$.
This is the only data that needs to be stored. While the algorithm
is described from the root and up, implementation is simpler (no recursion)
from the top and down to the root.

Often the source coder might be trained with many separate sequences,
rather than one long sequence. This is not an issue, but care has
to be taken with the startup for each sequence. The original CTW paper
\cite{WillemsAl95} assumes that a context of length at least $D$
is available prior to the start of the sequence, which is not true
in practice. The paper \cite{Willems98} solves this by introducing
an indeterminate context. A context may start with an indeterminate
context, but at most once. With multiple training sequences this could
happen more than once. A better approach is therefore to use the start
of each training sequence purely as a context (i.e., not code it).
This wastes some training bits, but if the sequences are long the
loss is minor. A different case would be if we need to find short
atypical sequences rather than subsequences. In that case a more careful
treatment of start of sequences would be needed.

\begin{figure}[tbh]
\begin{centering}
\includegraphics[width=3.5in]{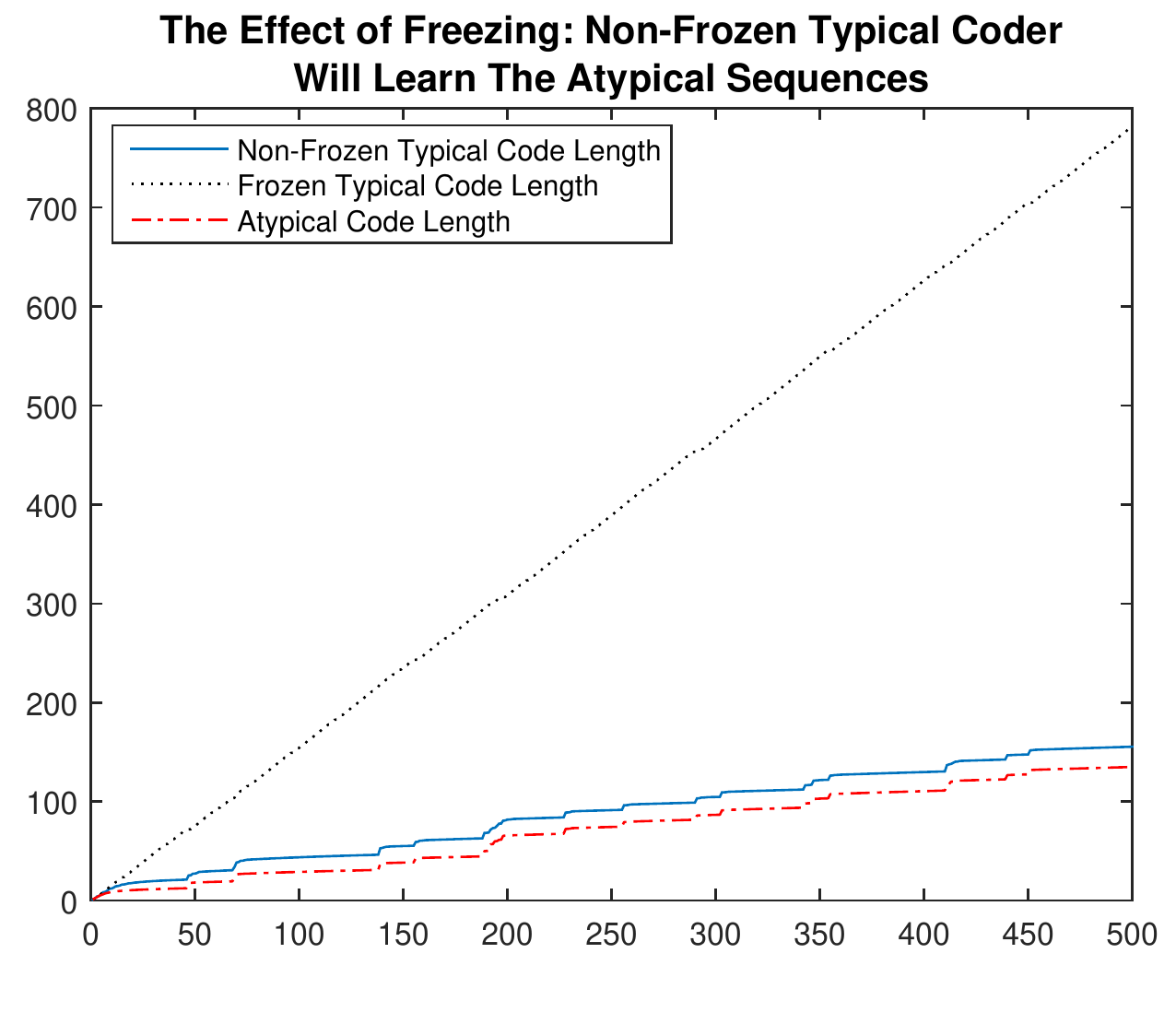} 
\par\end{centering}
\protect\caption{\label{Freezing.fig}The importance of freezing source coding when
testing for atypicality.}
\end{figure}

Freezing the encoder is essential in implementing atypicality. A simulation
confirming this is shown in Fig. \ref{Freezing.fig}. The CTW algorithm
is trained with a three-state Markov chain with transition probability
{[}.05 .95 0;0 .05 .95;.95 0 .05{]} while generating \{0,1\} according
to {[}0 1 x;x 1 0;1 x 0{]}, so this Markov chain mostly generates
the following pattern: {[}1 0 1{]}. The test sequence is generated
by another three-state Markov chain with the same transition probability
but generating \{0,1\} according to {[}0 1 x;x 1 0;0 x 1{]}, i.e.,
generating mostly {[}1 0 0{]} as pattern. With the non-frozen algorithm
the code length difference between typical and atypical encoding is
so small that it can easily be missed, although the difference in
the patterns themselves in the raw data is clearly visible to the
naked eye. The reason the non-frozen algorithm does not work is that
it quickly learns the new {[}1 0 0{]} pattern. Any good source coder
would do that including LZ. This is advantageous to source coding,
but in this case it means missing a very obvious atypical pattern.

If a very large amount of data is used for training, the complexity
can become very high, mainly in terms of memory. Namely, all contexts
might be observed, and the context tree completely filled out. For
example, suppose the typical data is actually iid. That means that
every string $x_{1},\ldots,x_{N}$ is seen with equal probability.
For the CTW algorithm that means that every node of the context tree
will be filled out, and the number of nodes with depth $D$ is $2^{D+1}$.
For dictionary based algorithms, it means that the dictionary size
becomes huge. What is needed is some algorithm that not only estimates
the unknown parameters, but also the model (e.g., iid). One way could
be to trim the context tree (or dictionary), but we have not looked
into this in detail.

\subsection{Atypical subsequences}

For finding atypical subsequences of a a long sequences, the same
basic setup as in the previous sections can be used. Let $\mathcal{X}(n,l)=(x_{n},x_{n+1},\ldots,x_{n+l-1})$
be a subsequence of $\{x_{n},n=0,\ldots,\infty\}$ that we want to
test for atypicality. As in Section \ref{Subsequences.sec} the start
of a sequence needs to be encoded as well as the length. Additionally
the code length is minimized over the maximum depth $D$ of the context
tree. The atypical code length is then given by 
\begin{align*}
L_{A}(\mathcal{X}(n,l))= & \min_{D}\left(-\log P_{w}^{\lambda}(D)+\log^{*}D\right)+\log^{*}l
\end{align*}
\emph{except} for the $\tau$. Here $P_{w}^{\lambda}(D)$ denotes
the probability at the root of the context tree \cite{WillemsAl95}
of depth $D$. Since we are also interested in finding short sequences,
how the encoding is initialized is of importance, and for atypical
coding we therefore use the algorithm in Section II of \cite{Willems98}. 

For typical coding we use either a known fixed model and Shannon codes,
or the algorithm in Section \ref{Typical.sec} when the model is not
known; when we encode $\mathcal{X}(n,l)$ we use $x_{n-D},\ldots x_{n-1}$
as context for $x_{n}$ (we can assume $n>D$). Equivalently, we can
encode the total sequence $\{x_{n},n=0,\ldots,\infty\}$ (with the
algorithm from Section \ref{Typical.sec}); let $L(n)$ be the codelength
for the sequence $x_{0},\ldots,x_{n}$. Then we can put $L_{T}(\mathcal{X}(n,l))=L(n+l-1)-L(n)$.

We need to test every subsequence of every length, that is, we need
to test subsequences $\mathcal{X}(n,l)$ for every value of $n$ and
$l$. For atypical coding this means that a new CTW algorithm needs
to be started at every sample time. So, if the maximum sequence length
is $L$, $L$ separate CTW trees need to maintained at any time. These
are completely independent, so they can be run on parallel processors.

The result is that for every bit of the data we calculate 
\begin{align}
\Delta L(n) & =L_{A}(\mathcal{X}(n,l))-\min_{l}L_{T}(\mathcal{X}(n,l))\label{DeltaLn.eq}
\end{align}
and we can state the atypicality criterion as $\Delta L(n)<-\tau$.
The advantage of stating it like this is that we do not need to choose
$\tau$ prior to running the algorithm. We can sort according to $\Delta L(n)$,
and first examine the data with smallest $\Delta L(n)$, which should
be the \emph{most} atypical parts of data. Thus, the algorithm is
really parameter free, which is one advantage of our approach. In
many anomaly detection algorithms, there are multiple parameters that
need to be adjusted.

This implementation clearly is quite complex, but still feasible to
implement for medium sized data sets due to the speed of the source
coder. In order to process larger data sets, a faster  approximate
search algorithm is needed, and we are working on such algorithms,
but leave that as a topic of later papers.

\section{Experimental Results}

In order to verify the performance of our algorithm, we used three
different experiments. In the first we evaluated randomness of sources,
in accordance with our starting point of Kolmogorov-Martin-L\"{o}f
randomness. In the second, we looked for infection in human DNA, and
in the third we looked for arrhythmia in ECG. 

A word about presentation of the results. For the outcome of our method
we plot $\Delta L(n)$ given by (\ref{DeltaLn.eq}). At the same time,
we would like to illustrate the raw data. The source in all cases
is a stream of bits $x[n]\in\{0,1\}$. We convert this to $y[n]\in\{-1,1\}$
(i.e, $y[n]=(-1)^{x[n]-1}$), and then plot $S[N]=\sum_{n=1}^{N}y[n]$;
we call this the random walk representation. In our experience, this
allows one to quickly assess if there is any obvious pattern in data.
If the data is random, the results will look like a typical random
walk: both small fluctuations and large fluctuations. 

All experimental data and software used is available at http://itdata.hostmadsen.com.

\subsection{\label{subsec:Coin-Tosses}Coin Tosses}

In this experiment the typical data is iid binary random. As source
of typical data we used experimental coin tosses from \cite{CoinToss}.
This data consists of 40,000 tosses by two Berkeley undergraduates
of a fair coin and the result has 20,217 heads ($X_{i}=1$), so $\Pr\{X_{i}=1\}=0.505425\approx\frac{1}{2}$.
Therefore we can consider it as a real binary IID experiment, indeed
it is an example of pure random data. In our experiments with this
data, we examine the randomness of other types of data. 

\begin{figure}[tbh]
\includegraphics[width=3.5in]{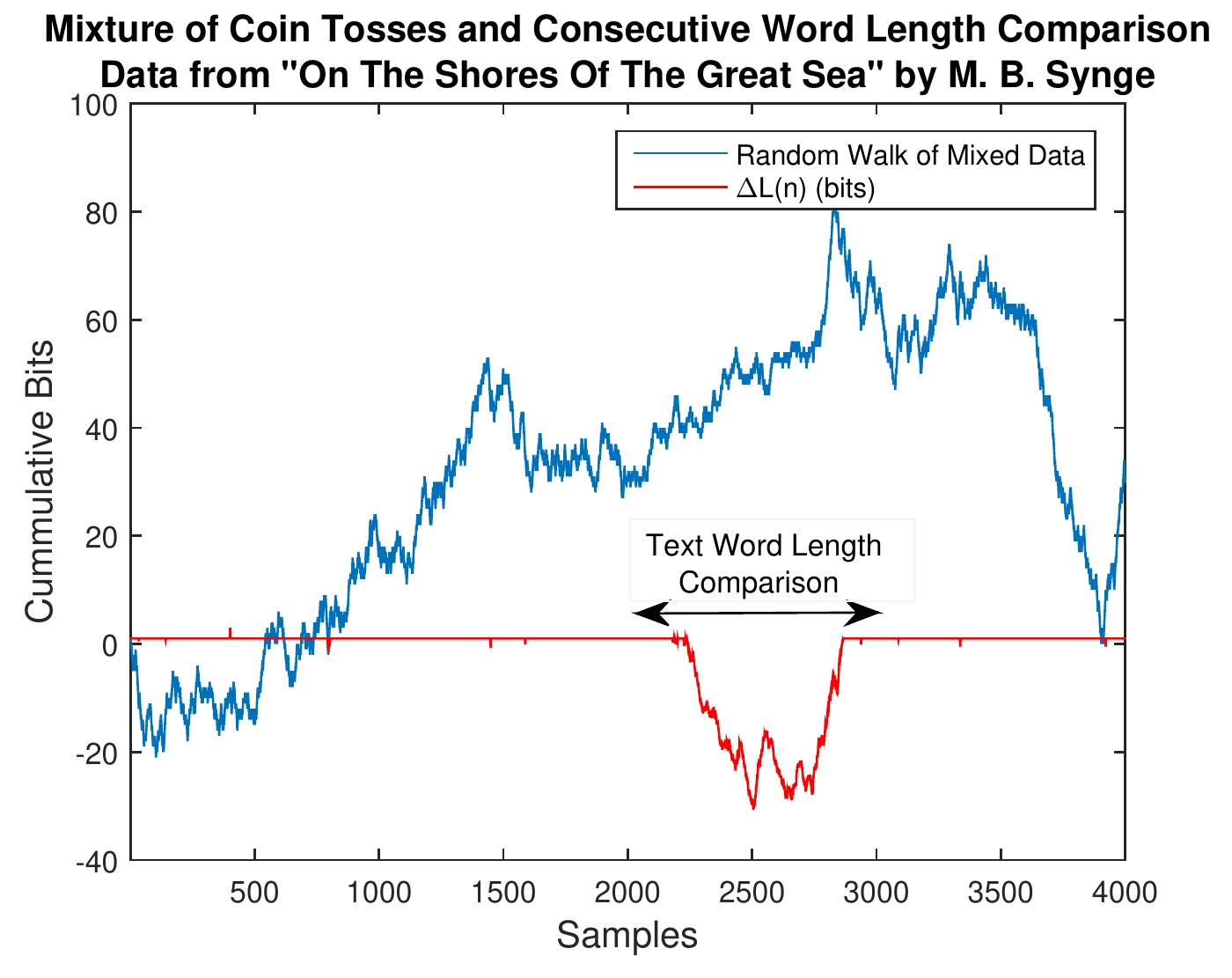}

\caption{\label{fig:CoinTossText}Random walk of mixed coin tosses and consecutive
word comparison.}
\end{figure}

One type of data one might \emph{think} is purely random are word
length changes in a text. In the first experiment, we generate binary
data using consecutive word length comparison of part of M. B. Synge's
``On the shores of the great sea'' in the following manner: If the
next word is longer than the current word, 1 is assigned to the binary
data, otherwise 0. In the case of two consecutive words with same
length, a random 0 or 1 is generated (with a \emph{good} random number
generator). We then insert this data in the coin toss data. Since
we assume coin tosses data is IID, there is no need to train the CTW
and the the code length of the IID case (\ref{TypicalCL.eq}) can
be used for typical coding. Fig. \ref{fig:CoinTossText} illustrates
the result of the algorithm on the mixed data. Thus, word length changes
are not iid random. Perhaps this is because word length is bounded
from above and below, so that there are limits to how long runs of
0s or 1s are possible. 

\begin{figure}[tbh]
\includegraphics[width=3.5in]{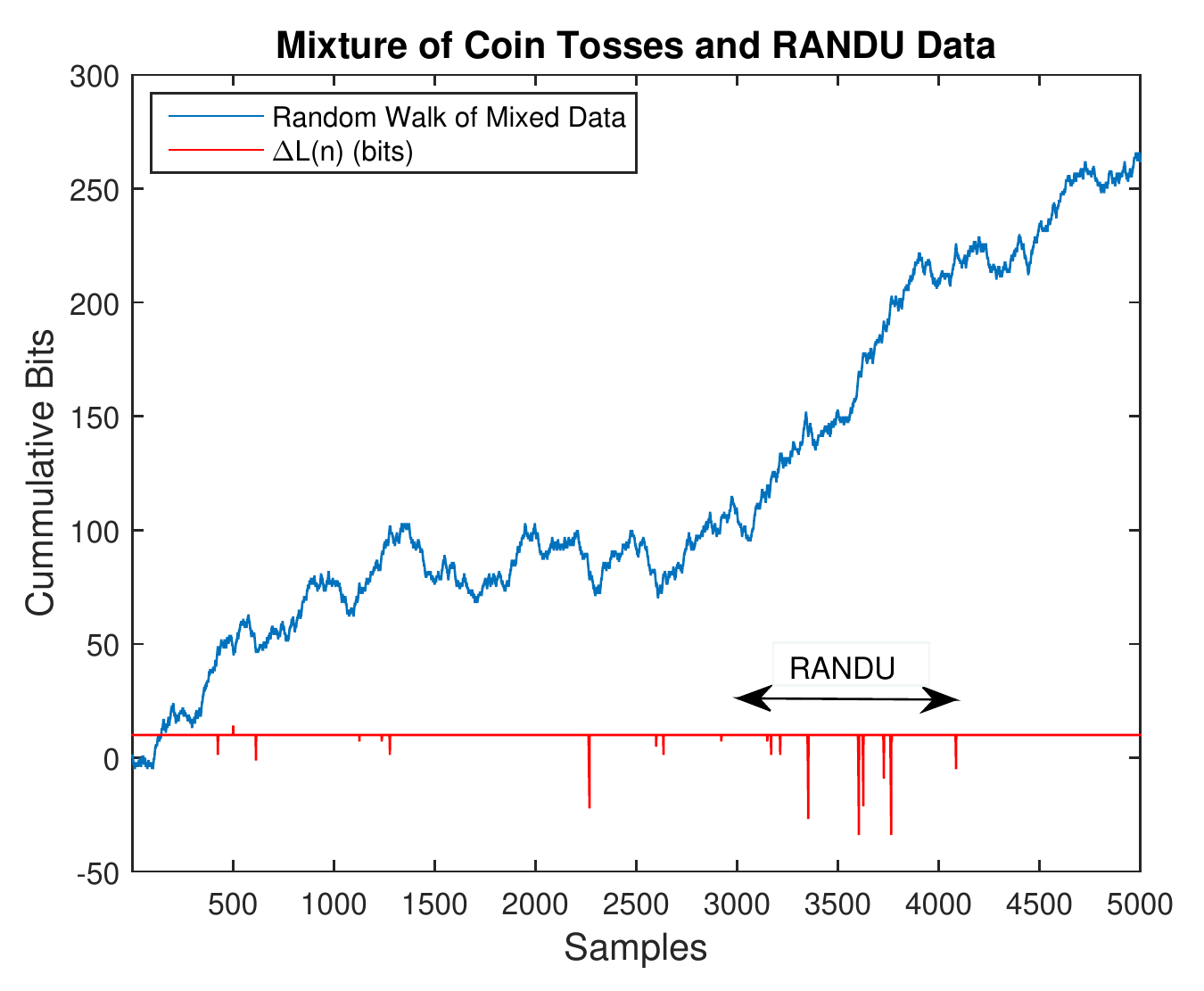}

\caption{\label{fig:CoinTossRandu}Random walk of mixed coin tosses and RANDU.}
\end{figure}

In second experiment, we generated random data with the infamous RANDU
random number generator. This was a random number generator that was
widely used until it was discovered that it has some clear deviation
from randomness. RANDU generates random numbers in the interval $[0,2^{31}-1]$,
so each number needs 31 bits for binary representation. But instead
of using 31 bits for each number, we sum up all the 31 bits and compare
it with 15.5 to generate either 0 or 1. Then this data is inserted
in the part of coin tosses data. Fig. \ref{fig:CoinTossRandu} shows
that the most atypical segment is where we have inserted data from
RANDU random number generator.

In the third experiment, we generate binary data using consecutive
heart rate comparison of part of normal sinus rhythm downloaded from
MIT- BIH database \cite{physionet} in the same way as for the text.
Fig. \ref{fig:CoinTossRandu} represents the result of the algorithm
on the mixed data. As can be seen the atypicality measure shows a
huge difference between iid randomness and randomness of consecutive
heart beats. We don't know why this is the case. 

\begin{figure}[tbh]
\includegraphics[width=3.5in]{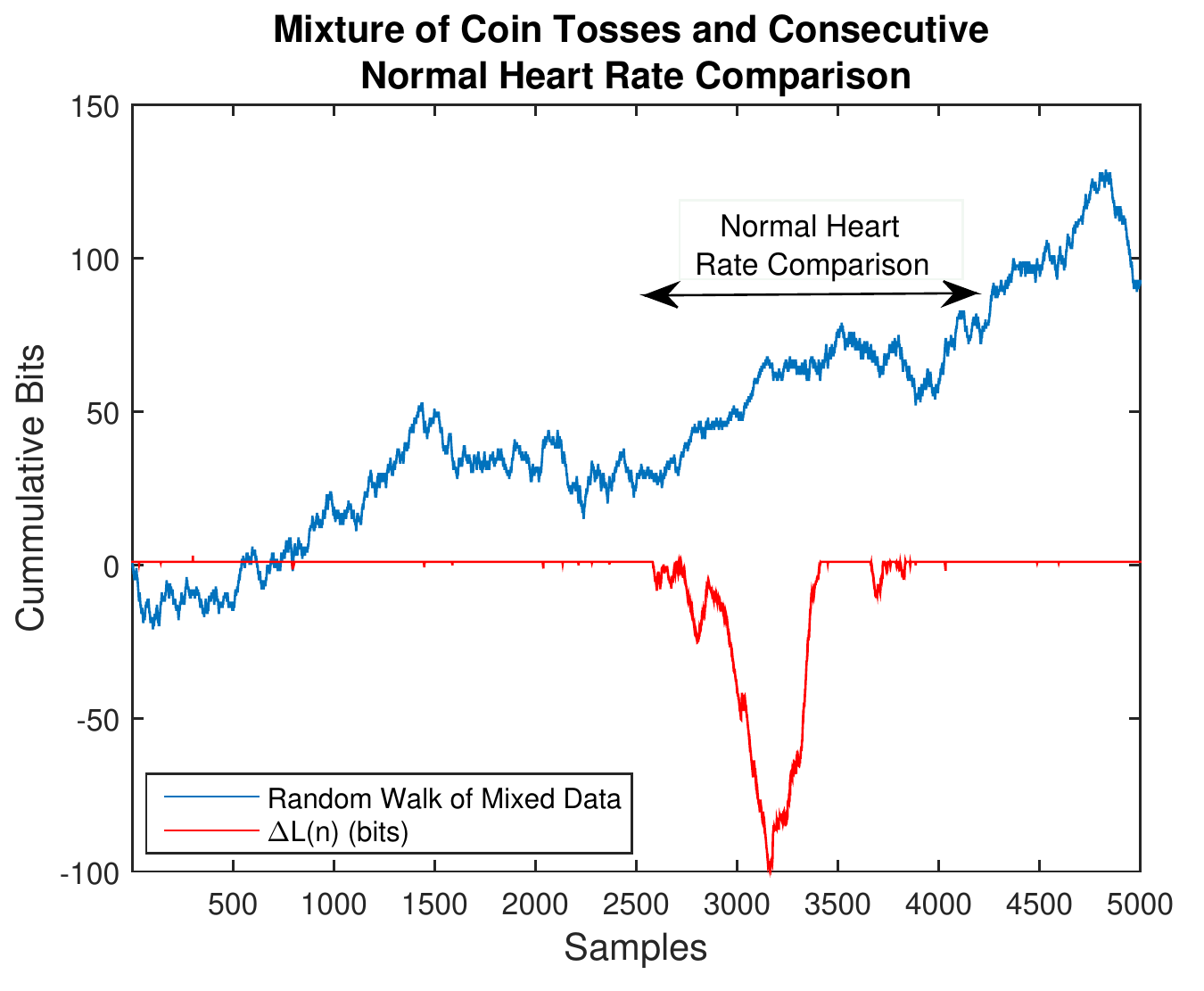}

\caption{\label{fig:CoinTossHrv}Random walk of mixed coin tosses and HRV.}
\end{figure}

\subsection{DNA}

In this collection of experiments, we detect viral and bacterial insertion
into human genomic DNA. DNA from foreign species can be inserted into
the human genome either through natural processes \cite{Britten03091996},
typically though viral infections, bacterial infections, or through
genetic engineering \cite{Zaman12}. The inverse also occurs in genetic
engineering experiments during the creation of \textquotedblleft transgenic\textquotedblright{}
organisms, with the insertion of human DNA into bacteria, yeast, worms,
or mice. In the experiments we show here, we have focused on the former
case. We train the CTW algorithm on pure human genomic DNA. 

The data that we have used was comprised of \textasciitilde{}20 kilobases
of human genomic DNA (each sequence from a different chromosome) with
either bacterial or viral random DNA sequences (\textasciitilde{}2
kilobases per insertion) inserted. Since our software is too slow
to find atypical sequences of length more than a few hundreds, we
removed the middle of the insertions. Notice that this actually makes
detection harder. We used some of the human DNA for training, but
not the same as the test sequences. 

\begin{figure}[tbh]
\includegraphics[width=3.5in]{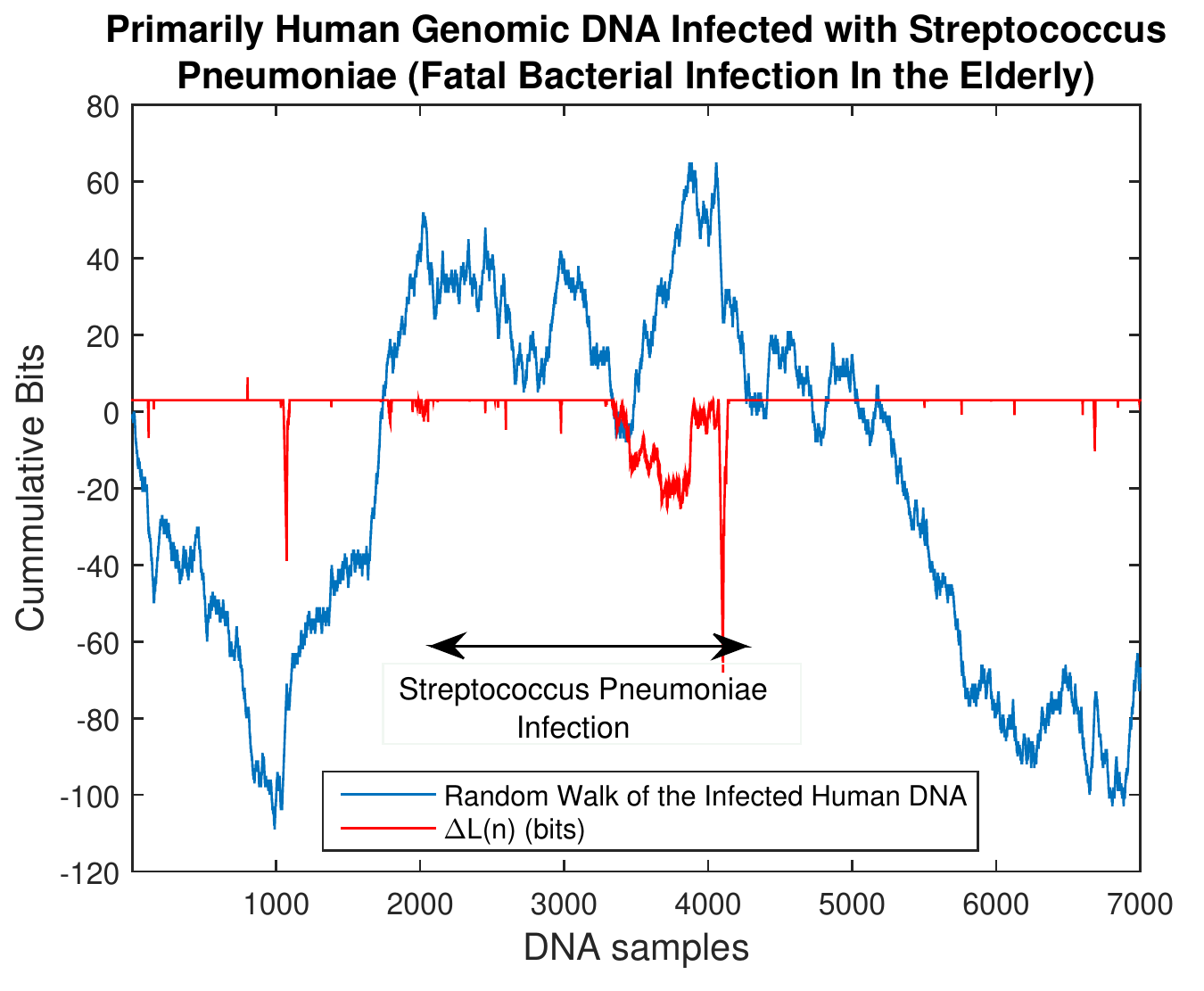}

\caption{\label{fig:InfectedBacterialDNA}Random walk of human DNA with bactrial
infection.}
\end{figure}

In the first experiment we tried to detect short sequences from Streptococcus
Pneumoniae (a bacterial infection with a high fatality rate, and a
frequent cause of death in the elderly) randomly inserted into larger
segments of human genomic DNA. Fig. \ref{fig:InfectedBacterialDNA}
illustrates the result of the experiment. Based on the figure, the
inserted Streptococcus Pneumoniae DNA fragment was detected by our
algorithms. 

\begin{figure}[tbh]
\includegraphics[width=3.5in]{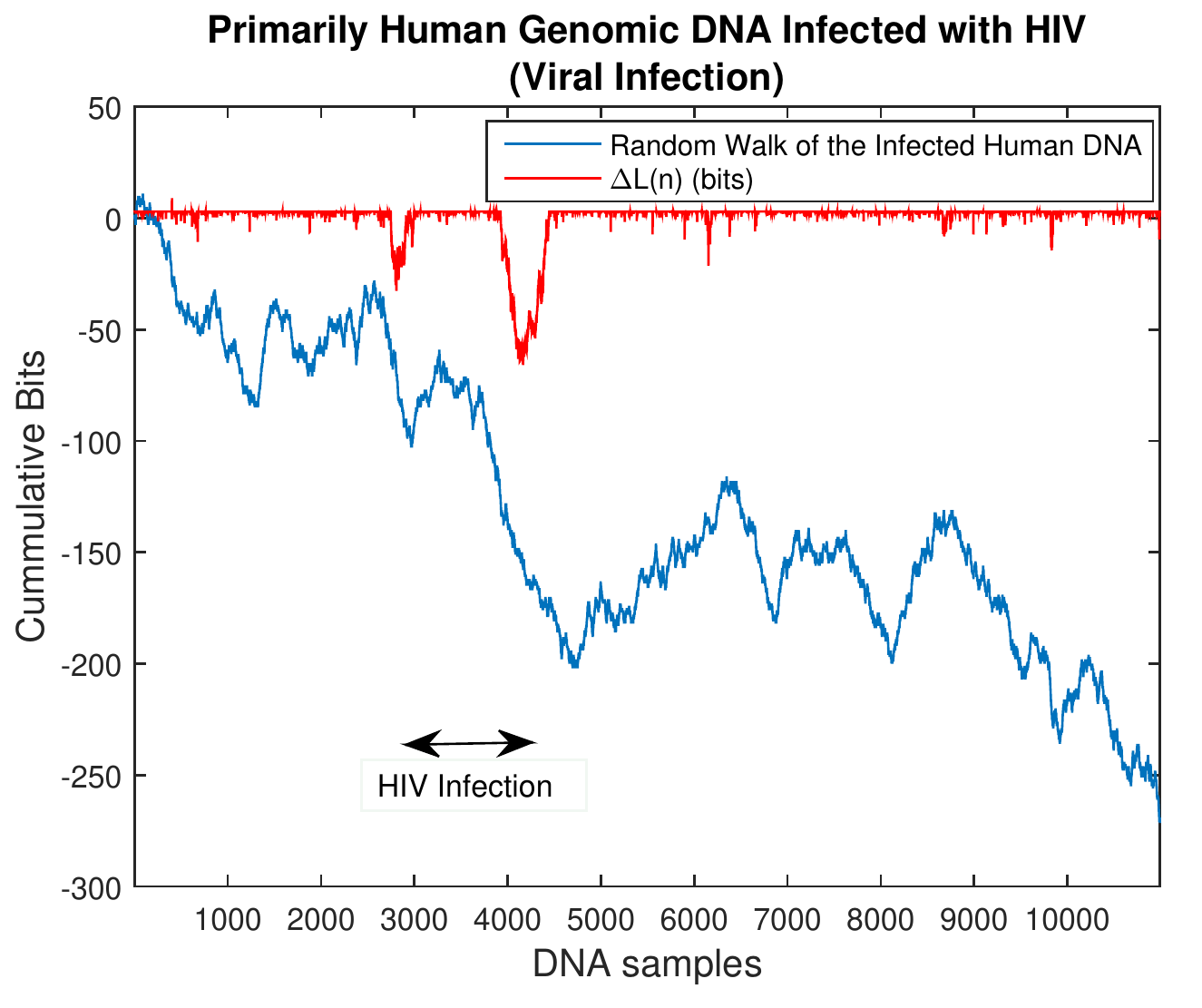}

\caption{\label{fig:InfectedViralDNA}Random walk of human DNA with viral infection.}
\end{figure}

In the second experiment we tried to detect HIV inserted into human
genomic DNA to mimic viral infection, which is a more realistic experiment
since viruses typically insert their DNA into the host genome every
time a human obtains a viral infection. Fig. \ref{fig:InfectedViralDNA}
illustrates the result of the experiment. As can be seen, the infected
viral fragment was detected by our algorithms. 

\subsection{HRV}

\begin{figure}[tbh]
\includegraphics[width=3.5in]{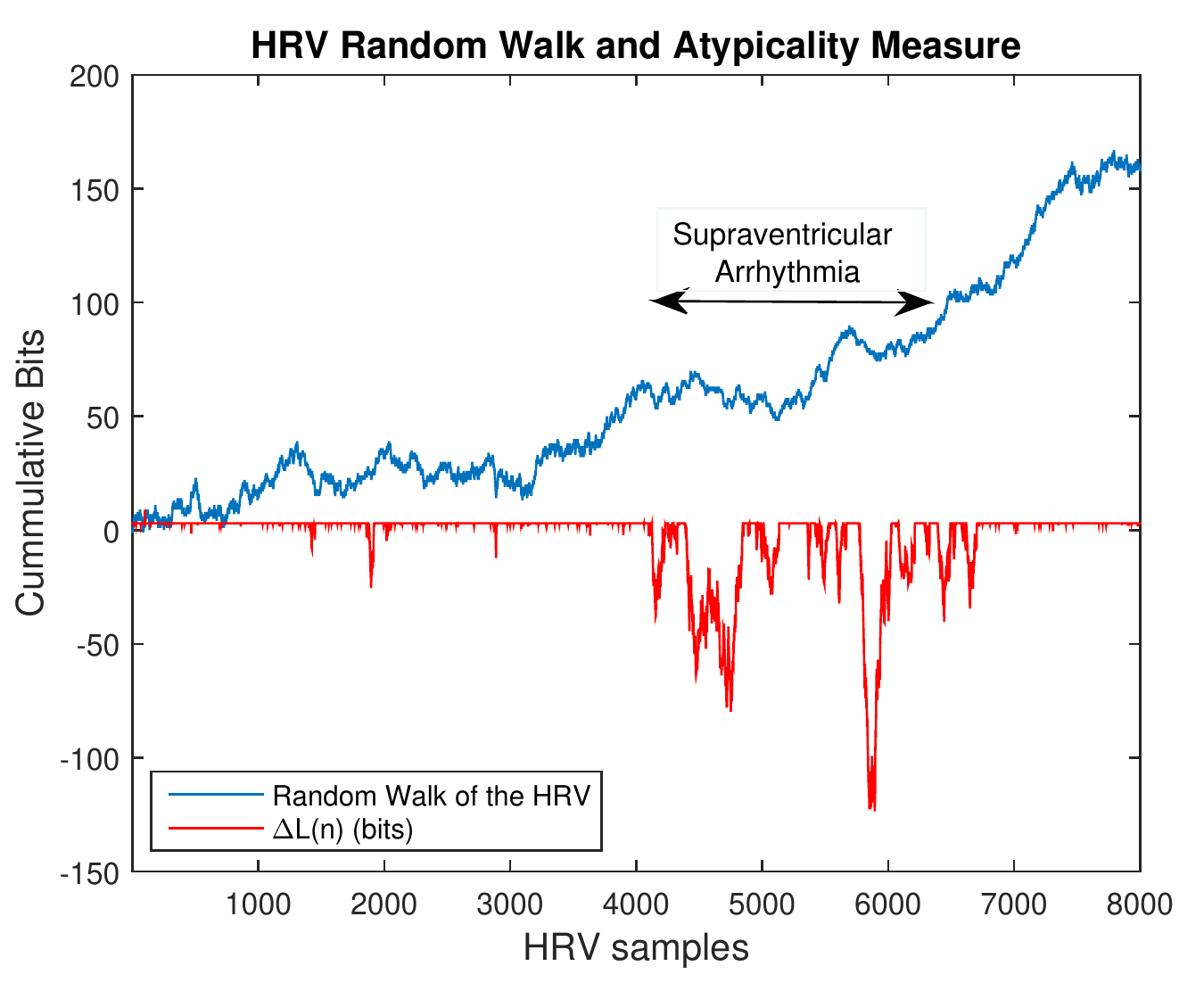}

\caption{\label{fig:HRV}Random walk of HRV.}
\end{figure}

While HRV (Heart Rate Variability) can be a powerful indicator for
arrhythmia \cite{Malik96}, the common issue is that it is not known
exactly what to look for in the data. Our aim for this application
is to use atypicality to localize signs of subtle and complex arrhythmia.
In \cite{Host13ITW} based on our modest goal of localizing a simple
type of known arrhythmia, we managed to find premature beats using
HRV signal, but here we attempted to detect more subtle arrhythmia.
The HRV signals that we used were downloaded from MIT- BIH database
\cite{physionet}. We used \textquotedblleft MIT-BIH normal sinus
rhythm database (nsrdb)\textquotedblright{} and \textquotedblleft MIT-BIH
supraventricular arrhythmia database (svdb)\textquotedblright . Encoding
of HRV signals were done by same manner as the text word length comparison
of subsection \ref{subsec:Coin-Tosses}. In this experiment, CTW was
trained with HRV of normal sinus rhythm, then applied to a HRV signal
that has supraventricular rhythms. Fig. \ref{fig:HRV} shows the result
of the simulation. The algorithm was able to localize the segment
that suffers from abnormal rhythms.

\section{Conclusion}

In this paper we have developed a criterion for finding atypical (sub)sequences
in large datasets. The criterion is based on a solid theoretical foundation,
and we have shown that the criterion is amenable to theoretical analysis.
In particular, we have shown that the probability that a sample is
intrinsically atypical is less than 1, an important theoretical requirement
that is not trivially satisfied.

We have also shows in a few examples with real-world data that the
method is able to find known atypical subsequences. Our aim set out
in the introduction was more ambitiously to find 'interesting' data
in large datasets. In that context, the purpose of the current paper
is only to introduce the methodology, and show some theoretical properties.
In order to analyze really big datasets, we need much faster (probably
approximate) algorithms, much more efficient software (say in Python
or C), and faster computers. We are working on that as future work.

\bibliographystyle{IEEEtran}
\bibliography{Coop06,Sensor,ahmref2,Coop03,BigData,Underwater,CDMA,combined,ECGandHRV,nosal}

\begin{thebibliography}{10}
\providecommand{\url}[1]{#1}
\csname url@samestyle\endcsname
\providecommand{\newblock}{\relax}
\providecommand{\bibinfo}[2]{#2}
\providecommand{\BIBentrySTDinterwordspacing}{\spaceskip=0pt\relax}
\providecommand{\BIBentryALTinterwordstretchfactor}{4}
\providecommand{\BIBentryALTinterwordspacing}{\spaceskip=\fontdimen2\font plus
\BIBentryALTinterwordstretchfactor\fontdimen3\font minus
  \fontdimen4\font\relax}
\providecommand{\BIBforeignlanguage}[2]{{%
\expandafter\ifx\csname l@#1\endcsname\relax
\typeout{** WARNING: IEEEtran.bst: No hyphenation pattern has been}%
\typeout{** loaded for the language `#1'. Using the pattern for}%
\typeout{** the default language instead.}%
\else
\language=\csname l@#1\endcsname
\fi
#2}}
\providecommand{\BIBdecl}{\relax}
\BIBdecl

\bibitem{AkogluAl12}
L.~Akoglu, H.~Tong, J.~Vreeken, and C.~Faloutsos, ``Fast and reliable anomaly
  detection in categorical data,'' in \emph{Proceedings of the 21st ACM
  international conference on Information and knowledge management}.\hskip 1em
  plus 0.5em minus 0.4em\relax ACM, 2012, pp. 415--424.

\bibitem{SmetsVreeken11}
\BIBentryALTinterwordspacing
K.~Smets and J.~Vreeken, \emph{The Odd One Out: Identifying and Characterising
  Anomalies}, 2011, ch.~69, pp. 804--815. [Online]. Available:
  \url{http://epubs.siam.org/doi/abs/10.1137/1.9781611972818.69}
\BIBentrySTDinterwordspacing

\bibitem{LiuPrincipe09}
W.~Liu, I.~Park, and J.~Principe, ``An information theoretic approach of
  designing sparse kernel adaptive filters,'' \emph{Neural Networks, IEEE
  Transactions on}, vol.~20, no.~12, pp. 1950--1961, Dec 2009.

\bibitem{ChandolaAl12}
V.~Chandola, A.~Banerjee, and V.~Kumar, ``Anomaly detection for discrete
  sequences: A survey,'' \emph{Knowledge and Data Engineering, IEEE
  Transactions on}, vol.~24, no.~5, pp. 823 --839, may 2012.

\bibitem{Rumsfeld}
D.~Rumsfeld, \emph{Known and Unknown: A Memoir}.\hskip 1em plus 0.5em minus
  0.4em\relax Penguin, 2011.

\bibitem{HamanoYamamoto08}
K.~Hamano and H.~Yamamoto, ``A randomness test based on t-codes,'' in
  \emph{Information Theory and Its Applications, 2008. ISITA 2008.
  International Symposium on}, dec. 2008, pp. 1 --6.

\bibitem{NiesBook}
A.~Nies, \emph{Computability and Randomness}.\hskip 1em plus 0.5em minus
  0.4em\relax Oxford University Press, 2009.

\bibitem{Malik96}
M.~Malik, ``Heart rate variability,'' \emph{Annals of Noninvasive
  Electrocardiology}, vol.~1, no.~2, pp. 151--181, April 1996.

\bibitem{M.F.Hilton:1999fj}
M.~F. Hilton, R.~A. Bates, K.~R. Godfrey, and et~al., ``Evaluation of frequency
  and time-frequency spectral analysis of heart rate variability as a
  diagnostic marker of the sleep apnea syndrome,'' \emph{Med. Biol. Eng.
  Comput.}, vol.~37, no.~6, pp. 760--769, November 1999.

\bibitem{N.V.Thakor:1991gf}
N.~V. Thakor and Y.~S. Zhu, ``Application of adaptive filtering to {ECG}
  analysis: Noise cancellation and arrhythmia detection,'' \emph{IEEE Trans. on
  Biomed. Eng.}, vol.~38, pp. 785--794, August 1991.

\bibitem{ThayerAl09}
J.~F. Thayer, S.~S. Yamamoto, and J.~F. Brosschot, ``The relationship of
  autonomic imbalance, heart rate variability and cardiovascular disease risk
  factors,'' \emph{International Journal of Cardiology}, vol. 141, no.~2, pp.
  122--131, May 2009.

\bibitem{FondonGarner04}
J.~Fondon and H.~Garner, ``Probing human cardiovascular congenital disease
  using transgenic mouse models,'' \emph{Proc Natl Acad Sci {U S A}}, vol. 101,
  no.~52, pp. 18\,058--63, 2004.

\bibitem{Mellinger07}
D.~K. Mellinger, K.~M. Stafford, S.~E. Moore, R.~P. Dziak, and H.~Matsumoto,
  ``An overview of fixed passive acoustic observation methods for cetaceans,''
  \emph{Oceanograpy}, vol.~20, no.~4, pp. 36--45, 2007.

\bibitem{ThottanJi03}
M.~Thottan and C.~Ji, ``Anomaly detection in ip networks,'' \emph{Signal
  Processing, IEEE Transactions on}, vol.~51, no.~8, pp. 2191 -- 2204, aug.
  2003.

\bibitem{LiVitanyi}
M.~Li and P.~Vit\'anyi, \emph{An Introduction to Kolmogorov Complexity and Its
  Applications}, 3rd~ed.\hskip 1em plus 0.5em minus 0.4em\relax Springer, 2008.

\bibitem{CoverBook}
T.~Cover and J.~Thomas, \emph{Information Theory, 2nd Edition}.\hskip 1em plus
  0.5em minus 0.4em\relax John Wiley, 2006.

\bibitem{Rissanen86b}
J.~Rissanen, ``Complexity of strings in the class of markov sources,''
  \emph{Information Theory, IEEE Transactions on}, vol.~32, no.~4, pp. 526 --
  532, jul 1986.

\bibitem{Rissanen83}
------, ``A universal prior for integers and estimation by minimum description
  length,'' \emph{The Annals of Statistics}, no.~2, pp. 416--431, 1983.

\bibitem{Rissanen84}
------, ``Universal coding, information, prediction, and estimation,''
  \emph{Information Theory, IEEE Transactions on}, vol.~30, no.~4, pp. 629 --
  636, jul 1984.

\bibitem{Rissanen86}
------, ``Stochastic complexity and modeling,'' \emph{The Annals of
  Statistics}, no.~3, pp. 1080--1100, Sep. 1986.

\bibitem{EvansBarnettAl04}
S.~Evans, B.~Barnett, S.~Bush, and G.~Saulnier, ``Minimum description length
  principles for detection and classification of ftp exploits,'' in
  \emph{Military Communications Conference, 2004. MILCOM 2004. 2004 IEEE},
  vol.~1, oct.-3 nov. 2004, pp. 473 -- 479 Vol. 1.

\bibitem{WangAl12}
N.~Wang, J.~Han, and J.~Fang, ``An anomaly detection algorithm based on
  lossless compression,'' in \emph{Networking, Architecture and Storage (NAS),
  2012 IEEE 7th International Conference on}, 2012, pp. 31--38.

\bibitem{LeeXiang01}
W.~Lee and D.~Xiang, ``Information-theoretic measures for anomaly detection,''
  in \emph{Security and Privacy, 2001. S P 2001. Proceedings. 2001 IEEE
  Symposium on}, 2001, pp. 130--143.

\bibitem{PaschalidisSmaragdakis09}
I.~Paschalidis and G.~Smaragdakis, ``Spatio-temporal network anomaly detection
  by assessing deviations of empirical measures,'' \emph{Networking, IEEE/ACM
  Transactions on}, vol.~17, no.~3, pp. 685--697, 2009.

\bibitem{HanChoi09}
C.-K. Han and H.-K. Choi, ``Effective discovery of attacks using entropy of
  packet dynamics,'' \emph{Network, IEEE}, vol.~23, no.~5, pp. 4--12, 2009.

\bibitem{BeligaLin05}
P.~Baliga and T.~Lin, ``Kolmogorov complexity based automata modeling for
  intrusion detection,'' in \emph{Granular Computing, 2005 IEEE International
  Conference on}, vol.~2, 2005, pp. 387--392 Vol. 2.

\bibitem{ShahriarZulkernine12}
H.~Shahriar and M.~Zulkernine, ``Information-theoretic detection of sql
  injection attacks,'' in \emph{High-Assurance Systems Engineering (HASE), 2012
  IEEE 14th International Symposium on}, 2012, pp. 40--47.

\bibitem{XiangLiZhou11}
Y.~Xiang, K.~Li, and W.~Zhou, ``Low-rate ddos attacks detection and traceback
  by using new information metrics,'' \emph{Information Forensics and Security,
  IEEE Transactions on}, vol.~6, no.~2, pp. 426--437, 2011.

\bibitem{PanWang06}
F.~Pan and W.~Wang, ``Anomaly detection based-on the regularity of normal
  behaviors,'' in \emph{Systems and Control in Aerospace and Astronautics,
  2006. ISSCAA 2006. 1st International Symposium on}, 2006, pp. 6 pp.--1046.

\bibitem{EllandLiebrock06}
E.~Eiland and L.~Liebrock, ``An application of information theory to intrusion
  detection,'' in \emph{Information Assurance, 2006. IWIA 2006. Fourth IEEE
  International Workshop on}, 2006, pp. 16 pp.--134.

\bibitem{LiVitanyiAl04}
M.~Li, X.~Chen, X.~Li, B.~Ma, and P.~Vitanyi, ``The similarity metric,''
  \emph{Information Theory, IEEE Transactions on}, vol.~50, no.~12, pp. 3250 --
  3264, dec. 2004.

\bibitem{KeoghAl04}
E.~Keogh, S.~Lonardi, and C.~A. Ratanamahatana, ``Towards parameter-free data
  mining,'' in \emph{Proceedings of the tenth ACM SIGKDD international
  conference on Knowledge discovery and data mining}.\hskip 1em plus 0.5em
  minus 0.4em\relax ACM, 2004, pp. 206--215.

\bibitem{KayBook}
S.~M. Kay, \emph{Fundamentals of Statistical Signal Processing, Volume I:
  Estimation Theory}.\hskip 1em plus 0.5em minus 0.4em\relax Prentice-Hall,
  1993.

\bibitem{ScharfBook}
L.~L. Scharf, \emph{Statistical Signal Processing: Detection, Estimation, and
  Time Series Analysis}.\hskip 1em plus 0.5em minus 0.4em\relax Addison-Wesley,
  1990.

\bibitem{Shamir06}
G.~Shamir, ``On the mdl principle for i.i.d. sources with large alphabets,''
  \emph{Information Theory, IEEE Transactions on}, vol.~52, no.~5, pp.
  1939--1955, May 2006.

\bibitem{Elias75}
P.~Elias, ``Universal codeword sets and representations of the integers,''
  \emph{Information Theory, IEEE Transactions on}, vol.~21, no.~2, pp. 194 --
  203, mar 1975.

\bibitem{PoorBook}
H.~V. Poor, \emph{An Introduction to Signal Detection and Estimation}.\hskip
  1em plus 0.5em minus 0.4em\relax Springer-Verlag, 1994.

\bibitem{hoeffding1963probability}
W.~Hoeffding, ``Probability inequalities for sums of bounded random
  variables,'' \emph{Journal of the American statistical association}, vol.~58,
  no. 301, pp. 13--30, 1963.

\bibitem{DemboBook}
A.~Dembo and O.~Zeitouni, \emph{Large Deviations Techniques and
  Applications}.\hskip 1em plus 0.5em minus 0.4em\relax Springer, 1998.

\bibitem{LehmannBook}
E.~L. Lehmann, \emph{Testing Statistical Hypotheses}.\hskip 1em plus 0.5em
  minus 0.4em\relax Springer, 2005.

\bibitem{GrimmettBook}
G.~R. Grimmett and D.~R. Stirzaker, \emph{Probability and Random Processes,
  Third Edition}.\hskip 1em plus 0.5em minus 0.4em\relax Oxford University
  Press, 2001.

\bibitem{Rissanen78}
J.~Rissanen, ``Modeling by shortest data description,'' \emph{Automatica}, pp.
  465--471, 1978.

\bibitem{WillemsAl95}
F.~M.~J. Willems, Y.~Shtarkov, and T.~Tjalkens, ``The context-tree weighting
  method: basic properties,'' \emph{Information Theory, IEEE Transactions on},
  vol.~41, no.~3, pp. 653--664, 1995.

\bibitem{ZivLempel77}
J.~Ziv and A.~Lempel, ``A universal algorithm for sequential data
  compression,'' \emph{Information Theory, IEEE Transactions on}, vol.~23,
  no.~3, pp. 337 -- 343, may 1977.

\bibitem{ZivLempel78}
------, ``Compression of individual sequences via variable-rate coding,''
  \emph{Information Theory, IEEE Transactions on}, vol.~24, no.~5, pp. 530 --
  536, sep 1978.

\bibitem{EffrosVerduAl02}
M.~Effros, K.~Visweswariah, S.~Kulkarni, and S.~Verdu, ``Universal lossless
  source coding with the burrows wheeler transform,'' \emph{Information Theory,
  IEEE Transactions on}, vol.~48, no.~5, pp. 1061 --1081, may 2002.

\bibitem{ClearyWitten84}
J.~Cleary and I.~Witten, ``Data compression using adaptive coding and partial
  string matching,'' \emph{Communications, IEEE Transactions on}, vol.~32,
  no.~4, pp. 396 -- 402, apr 1984.

\bibitem{Moffat90}
A.~Moffat, ``Implementing the ppm data compression scheme,''
  \emph{Communications, IEEE Transactions on}, vol.~38, no.~11, pp. 1917
  --1921, nov 1990.

\bibitem{Titchener98}
M.~Titchener, ``Deterministic computation of complexity, information and
  entropy,'' in \emph{Information Theory, 1998. Proceedings. 1998 IEEE
  International Symposium on}, aug 1998, p. 326.

\bibitem{KawaharadaSpeidelAl05}
K.~Kawaharada, K.~Ohzeki, and U.~Speidel, ``Information and entropy
  measurements on video sequences,'' in \emph{Information, Communications and
  Signal Processing, 2005 Fifth International Conference on}, 0-0 2005, pp.
  1150 --1154.

\bibitem{Speidel09}
U.~Speidel, ``A note on the estimation of string complexity for short
  strings,'' in \emph{Information, Communications and Signal Processing, 2009.
  ICICS 2009. 7th International Conference on}, dec. 2009, pp. 1 --5.

\bibitem{SpeidelAl07}
U.~Speidel, R.~Eimann, and N.~Brownlee, ``Detecting network events via
  t-entropy,'' in \emph{Information, Communications Signal Processing, 2007 6th
  International Conference on}, dec. 2007, pp. 1 --5.

\bibitem{SpeidelGulliver12}
U.~Speidel and T.~Gulliver, ``An analytic upper bound on t-complexity,'' in
  \emph{Information Theory Proceedings (ISIT), 2012 IEEE International
  Symposium on}, july 2012, pp. 2706 --2710.

\bibitem{YangSpeidel05}
J.~Yang and U.~Speidel, ``String parsing-based similarity detection,'' in
  \emph{Information Theory Workshop, 2005 IEEE}, aug.-1 sept. 2005, p. 5 pp.

\bibitem{VolfWillems98}
P.~A.~J. Volf and F.~M.~J. Willems, ``Switching between two universal source
  coding algorithms,'' in \emph{Data Compression Conference, 1998. DCC '98.
  Proceedings}, 1998, pp. 491--500.

\bibitem{JacquetSzpankowski11}
P.~Jacquet and W.~Szpankowski, ``Limiting distribution of lempel ziv'78
  redundancy,'' in \emph{Information Theory Proceedings (ISIT), 2011 IEEE
  International Symposium on}, 31 2011-aug. 5 2011, pp. 1509 --1513.

\bibitem{WillemsAl97}
F.~Willems, Y.~Shtarkov, and T.~Tjalkens, ``Reflections on "the context tree
  weighting method: Basic properties",'' \emph{Newsletter of the IEEE
  Information Theory Society}, vol.~47, no.~1, 1997.

\bibitem{Willems98}
F.~Willems, ``The context-tree weighting method: extensions,''
  \emph{Information Theory, IEEE Transactions on}, vol.~44, no.~2, pp.
  792--798, Mar 1998.

\bibitem{CoinToss}
``{Department of Statistics, UC Berkeley},'' {http://www.stat.berkeley.edu/}.

\bibitem{physionet}
``{PhysioBank ATM},'' {http://physionet.org/cgi-bin/atm/ATM/}.

\bibitem{Britten03091996}
\BIBentryALTinterwordspacing
R.~J. Britten, ``{DNA} sequence insertion and evolutionary variation in gene
  regulation,'' \emph{Proceedings of the National Academy of Sciences},
  vol.~93, no.~18, pp. 9374--9377, 1996. [Online]. Available:
  \url{http://www.pnas.org/content/93/18/9374.abstract}
\BIBentrySTDinterwordspacing

\bibitem{Zaman12}
J.~Z.~K. Khattak, S.~Rauf, Z.~Anwar, H.~M. Wahedi, and T.~Jamil, ``Recent
  advances in genetic engineering - a review,'' \emph{Current Research Journal
  of Biological Sciences}, vol.~4, no.~1, 2012.

\bibitem{Host13ITW}
A.~H{\o}st-Madsen, E.~Sabeti, and C.~Walton, ``Information theory for atypical
  sequences,'' in \emph{IEEE Information Theory Workshop (ITW'13), Seville,
  Spain}, 2013.

\end{thebibliography}

\end{document}